\documentclass[a4paper,fleqn,usenatbib]{mnras}
\usepackage[T1]{fontenc}
\usepackage{ae,aecompl}

\usepackage{graphicx}	
\usepackage{amsmath}	
\usepackage{amssymb}	
\usepackage{subfig}
\usepackage{hyperref}

\usepackage{adjustbox}
\usepackage{color}
\usepackage{upgreek}
\usepackage{float}

\usepackage{booktabs}

\title[Multi-wavelength study of BL\,Lac]{Multi-wavelength variability and broadband SED modeling of BL Lac during a bright flaring period MJD 59000--59943}

\author[Shah et al.]{
Zahir Shah$^1$\thanks{E-mail: shahzahir4@gmail.com} 
\\
$^{1}$ Department of Physics, Central University of Kashmir,  Ganderbal-191131, India}

\date{Accepted XXX. Received YYY; in original form ZZZ}

\pubyear{2023}

\begin{document}
\label{firstpage}
\pagerange{\pageref{firstpage}--\pageref{lastpage}}
\maketitle

\begin{abstract}
We carried a detailed temporal and spectral  study of  the  BL\,Lac  by using the long-term \emph{Fermi}-LAT and \emph{Swift}-XRT/UVOT observations, during the period MJD\,59000-59943. The daily-binned $\gamma$-ray light curve displays a maximum flux of $1.74\pm0.09\times 10^{-5} \rm ph\,cm^{-2}\,s^{-1}$  on MJD\,59868, which is the highest  daily $\gamma$-ray flux observed from BL\,Lac. The $\gamma$-ray variability is characterised  by power-spectral-density (PSD), r.m.s-flux relation and flux-distribution study.  We find that power-law  model fits the PSD  with  index $\sim 1$, which suggest for long memory process at work.  The  observed r.m.s.-flux relation exhibits a linear trend, which indicates that the $\gamma$-ray flux distribution follows a log-normal distribution.  The skewness/Anderson-Darling test and histogram-fit reject the normality of flux distribution, and instead suggest that the flux distribution is log-normal distribution. The fractional-variability amplitude shows that source is more variable in X-ray band  than in optical/UV/$\gamma$-ray bands. In order to obtain an insight into the underlying physical process,  we extracted broadband spectra from  different time periods of the lightcurve. The broadband spectra are statistically fitted with the convolved one-zone leptonic model with  different forms of the particle energy distribution. We found that spectral energy distribution during different flux states can be reproduced well with the synchrotron, synchrotron-self-Compton and external-Compton emissions from a broken power-law electron distribution, ensuring equipartition condition.  A comparison between the best fit physical parameters  show that the variation in different flux-states are mostly related to increase in the bulk-Lorentz factor and spectral hardening of the particle distribution.
\end{abstract}



\section{Introduction}
Blazars are extreme class of extragalactic sources  with powerful relativistic jet pointing close to the line of sight of observer \citep{1995PASP..107..803U}.   The  relativistic jet of non-thermal plasma originates near to the supermassive black hole and is launched to kiloparsec/mega parsec scales  (see e.g. \citet{2019ARA&A..57..467B}).  The close pointing of  relativistic jet produces  Doppler boosted  emission, which results in extreme observational properties.
Blazars show variability in flux and spectrum across the electromagnetic band,  the timescales of variability can go down to  few minutes  \citep[ see e.g,][]{2007ApJ...664L..71A, 2016ApJ...824L..20A} to years. The detailed analysis  of these observational features often help us to obtain insight into the underlying physics and structure of the emission region. For example, the short timescale variability in flux hints the emission occurs in the inner region of  jet near to the SMBH,  such regions are difficult to explore otherwise with the present imaging telescopes.
Blazars are broadly divided into two classes:  BL\,Lac objects and Flat Spectrum Radio Quasars (FSRQ). This divide   is usually based on  the characteristics of their optical spectrum.  BL\,Lac objects have weak or absent emission line features with equivalent width ($\rm EW< 5$ \AA), while FSRQs show prominent emission line features with $\rm EW > 5$ \AA\citep{1995PASP..107..803U}.

The broadband spectral energy distribution (SED) of blazars exhibit a characteristic double peaked spectrum.
The low energy component  peaking at optical/UV/X-ray energies is due to Doppler boosted synchrotron emission. While the high energy component which peaks at $\gamma$-ray energy is  explained in different scenarios. The most commonly considered scenario is the inverse-Compton (IC) process,  where either the synchrotron photons or external photons or both  gets up-scattered to high energies by the relativistic particles in the jet. In case the seed photons for the IC scattering are synchrotron photons, then the process is known as  synchrotron self Compton \citep[SSC:][]{1974ApJ...188..353J,1992ApJ...397L...5M, 1993ApJ...407...65G}, if the  photons are external to jet then the process is known as  external Compton \citep[EC:][]{1992A&A...256L..27D,1994ApJ...421..153S, 2000ApJ...545..107B, 2017MNRAS.470.3283S}. Another possible explanation for the high energy component is through the proton-synchrotron process or proton-photon interactions \citep{1992A&A...253L..21M, 1993A&A...269...67M, 2001APh....15..121M, 2003APh....18..593M}. These interactions are likely responsible for producing not only $\gamma$-ray photons but also high energy neutrinos. In fact,  TXS 0506+056 was the first flaring blazar to be associated with the high energy neutrino
IC-170922A \citep{2018Sci...361.1378I}. Since the detection of neutrino event from blazars, the hybrid models i.e., lepto-hadronic models have become more attractive for modeling the broadband SED of blazars  \citep[e.g.][]{2018ApJ...863L..10A, 2018ApJ...866..109S, 2018ApJ...864...84K, 2022MNRAS.509.2102G}.
Blazars are further classified based on  peak frequency of the  synchrotron component  ($\nu^p_{syn}$) into  low synchrotron peaked blazars (LSP; $\rm \nu^p_{syn} \leq 10^{14} $ Hz), intermediate synchrotron peaked blazars (ISP; $\rm 10^{14} \leq \nu^p_{syn} \leq 10^{15} $ Hz), and high synchrotron peaked blazars (HSP; $\rm \nu^p_{syn} \geq 10^{15} $ Hz) \citep{2010ApJ...716...30A} .

BL\,Lacertae (BL\,Lac) is a blazar located at a redshift of $z\sim 0.069$ \citep{1978ApJ...219L..85M}. The source is generally classified as  LSP  \citep{2018A&A...620A.185N},  but has also  been identified as ISP  \citep[see eg.][]{2011ApJ...741...30A, 2016A&A...592A..22H}. BL\,Lac  falls in the category of  TeV detected blazars \citep{2001ARep...45..249N, 2019A&A...623A.175M},  first VHE emission detected from source was reported in \citet{2001ARep...45..249N}. 
The sources had been reported in enhanced flux states over different energy bands by number of Astronomy telegrams  \citep[some  recent  telegrams are][]{2021ATel14777....1B, 2021ATel14826....1B, 2022ATel15688....1L, 2022ATel15730....1P}. These made it a  target of number of multi-wavelength observational studies  \citep[see e.g][]{2015MNRAS.452.4263G, 2015A&A...573A..69W, 2018ApJ...856...95A, 2019A&A...623A.175M,  2021MNRAS.507.5602P, 2022MNRAS.513.4645S} and gave us broader understanding of its emission properties. 
Moreover, a wide range of  interpretations have been used to explain the  variable emission of the source in different flux states. For example,  the low-activity state emission from the source with moderate correlated variability between UV and X-rays bands  are explained by the SSC and EC processes  \citep{2011ApJ...730..101A}.  A broadband SED modelling performed on the sources  in the low and high flux states showed that the two emission regions located at two different sites are required to explain the high energy emission in the source \citep{2021MNRAS.507.5602P}. These studies suggest that the broadband SEDs of BL\,Lac object  cannot be well described by a one-zone SSC model, and instead  complex models involving the multi-zone SSC or EC components are needed.
 \citet{2013MNRAS.436.1530R} explained  the variable emission from the source due to change in orientation of the emitting regions, while  \citet{2016ApJ...816...53W} used an extended data and explained the flaring emission in the source as  turbulent plasma flowing across quasi-stationary shocks.  Using the longterm optical photometric and polarimetric data of BL\,Lac, \citet{2009A&A...503..103B}  showed that the observed variability of BL\,Lac can be explained with  a steady component of high degree of polarisation $\sim 40\%$.

\citet{2009A&A...503..797G}  used the X-ray light curve of BL\,Lac to investigate  the nature of its variability. They showed that the X-ray light curve of BL\,Lac follows the  lognormal distribution, infact BL\,Lac was the first blazar in which lognormality was detected  \citep{2009A&A...503..797G}. The X-ray light curves used in their work were less variable compared to other blazars.  Moreover, the amplitude of variability was found to be proportional to the flux level.  The lognormal variability and linear r.m.s-flux relation in  the X-ray light curves had been observed  in various compact systems, such as Seyfert galaxies and X-ray binaries.  In these sources,  the variability in the emission  is  a result of fluctuations in the accretion disk \citep{2005MNRAS.359..345U,  2008bves.confE..14M},  these fluctuation propagate inwards and produces a multiplicative emission. This implies that the lognormal distribution of flux is being powered by multiplicative processes rather than additive. In case of blazars, the observation of lognormal distribution possibly imply  the disc variability is  imprinted on the jet, hence provides a possible link between  the accretion and jet properties in a blazar.

Number of multiwavelength works have been carried out to understand  the possibly physical scenario responsible for the variable emission \citep{2015MNRAS.452.4263G, 2015A&A...573A..69W, 2018ApJ...856...95A, 2019A&A...623A.175M,  2021MNRAS.507.5602P, 2022MNRAS.513.4645S}. These studies suggest  complex physical mechanisms are required for the emission in low and high flux states. In this work, we conducted a comprehensive multiwavelength study of BL\,Lac by utilizing more than two and half years of \emph{Fermi} data, which includes the most recent and brightest  $\gamma$-ray flare ever detected from the source. The brightest $\gamma$-ray flare has not been studied in previous works.    We conducted a comprehensive analysis of the $\gamma$-ray flux distribution of the BL\,Lac source, which has not been explored in earlier works.    We also developed a convolved one zone leptonic model and  incorporated it as a local convolution  model in XSPEC.  The model  provides the statistical  broadband fit to the observed SED for any input particle distribution. In this work, using the convolved one-zone leptonic,  underlying physical parameters responsible for the variations in different states are statistically constrained in BL\,Lac for the first time. Moreover, we constrained the underlying particle distribution responsible for the broadband emission using the $\chi^2$ test. 
The multiwavelength data used in this work is acquired from \emph{Fermi}-LAT, \emph{Swift}-XRT and \emph{Swift}-UVOT. The framework of this paper is as following:- the details of the multiwavelength data and the data analysis procedure are given in section \S\ref{sec:data_ana}. We presents the results of multiwavelength temporal and spectral analysis in section \S\ref{sec:results}. We summarise and discuss the results in section \S\ref{sec:discus}. A cosmology with $\rm \Omega_M = 0.3$, $\Omega_\Lambda = 0.7$ and $\rm H_0 = 71\rm km s^{-1} Mpc^{-1}$ is used in this work.

\section{Data Analysis}\label{sec:data_ana}
In order to examine the temporal and spectral properties of BL\,Lac, we first obtained mutiwavelenth data including Optical/UV, X-ray and $\gamma$-ray. The data is acquired from \emph{Swift}-UVOT,  \emph{Swift}-XRT and  \emph{Fermi}-LAT. Details of the observations and the data analysis procedure followed in our works are given below.

\subsection{\emph{Fermi}-LAT}
Fermi-LAT is a high energy space based telescope with wide field of view $\sim$ 2.3 Sr. It is one of  the two instrument onboard Fermi Gamma-ray Space Telescope (formerly called GLAST), which was launched by NASA in 2008. Fermi-LAT principally operates in scanning mode,  it  surveys the entire sky in the energy range $\sim$ 20 MeV- 500 GeV  every three hours \citep{2009ApJ...697.1071A}. 
In this work,  $\gamma$-ray data of BL\,Lac is  retrieved from \emph{Fermi}-LAT for the time period MJD\,59000-59943. The data is converted to science products using the \emph{Fermitools} (formally Science Tools) with version 2.2.0,  hosted on an Anaconda Cloud channel that is maintained by the Fermi Science Support Center  (FSSC).  We  followed  the standard analysis procedure described in the \emph{Fermi}-LAT documentation\footnote{http://fermi.gsfc.nasa.gov/ssc/data/analysis/} for the data reduction. The \emph{P8R3} events were extracted from the 15 degree ROI centred at the source location.  The events having high probability of being photons are included in  analysis by using the \emph{SOURCE} class events as ``\emph{evclass}=128, \emph{evtype}=3". Since  Earth limb is strong source of background $\gamma$-rays, we avoided  the contamination of $\gamma$-rays from the bright Earth limb by using a zenith angle cut of 90 degree to the data. This zenith cut is recommended by the LAT instrument team above 100 MeV.
 The good time intervals (gti) in which the satellite was working in standard data taking mode are selected by using a filter expression $(DATA_-QUAL>0) \&\&(LAT_-CONFIG==1)$. 
 We modelled the  Galactic diffuse emission component and the isotropic emission components  with \emph{$gll_-iem_-v07.fits$} and \emph{$iso_-P8R3_-CLEAN_-V3_-v1.txt$}, respectively, and the post launch instrument response function used is \emph{$P8R3_-SOURCE_-V3$}. 
All the sources in the 4FGL catalog  within  (15+10) degree ROI centred at the source location were included in the XML model file.   We initially carried the likelihood analysis for the entire period of data, while keeping  the spectral parameters of sources free within 15 degree ROI,  and freezing  the parameters of sources lying outside the ROI. We allowed to vary the photon index and normalisation of the Galactic diffuse component, and normalisation of isotropic component during the spectral fitting. In the output model file, we freezed the spectral parameters for  the background sources with test statistic,  TS $< 25$, and the output model  is finally used for the generation of light curve and the spectrum of the source. We considered the detection of the source only if TS > 9 \citep[$\sim$ 3 sigma detection;][]{1996ApJ...461..396M}.

\subsection{\emph{Swift}-XRT}
In this work, the X-ray data is acquired from the \emph{Swift}-XRT telescope onboard the \emph{Neil Gehrels Swift Observatory} \citep[]{2004ApJ...611.1005G}.  During the period MJD 59000--59943, a total of 100 \emph{Swift} observations of BL\,Lac source were available. 
We obtained the \emph{Swift}-XRT light curve such that each observation ID corresponds to one point in the X-ray light curve. The X-ray data is processed with the \emph{HEASoft} package (v6.30) and CALDB (v20220803). 
 We used the \emph{XRTDAS} software  and the Standard xrtpipeline (v0.13.7)  to create the cleaned event files. 
We run the  \emph{xrtpipeline} by giving the appropriate inputs to the task  following the \emph{Swift} X-ray data analysis thread page. The source and background regions are chosen by using the \emph{xrtgrblc} task \citep{2013ApJS..207...28S}. The task selects the source and background regions based on count rate such that a circular source region is selected for  count rate $\leq 0.5\,\rm counts\,s^{-1}$ and annular source  region is selected for count rate  $>0.5\,\rm counts\,s^{-1}$. While background region is chosen as annular region in all cases. 
Further, for the spectral analysis  we have used an automated online X-ray analysis tool available at the UK Swift Science Data Centre \citep{2009MNRAS.397.1177E}. This online tool gives the source, background, and ancillary response files necessary  for the  spectral analysis. We used the \emph{GRPPHA} task to rebin the source spectrum such that resultant spectrum has 20 counts per bin. The grouping is needed in order to evaluate the model based on the C-statistic (cstat).
The X-ray spectral analysis in the energy range 0.3--10 keV was performed using the \emph{XSPEC} package \citep{1996ASPC..101...17A} available with the \emph{HEASoft}. Since the value of  Galactic neutral hydrogen column density ($\rm n_H$) had been reported to vary between $(1.7-2.8)\times 10^{21} \rm cm^{-2}$ \citep{2020ApJ...900..137W}, therefore we fitted  0.3--10 keV spectra  with an absorbed power-law (PL) by choosing the $\rm n_H$ value between  $(1.7-2.8)\times 10^{21}~\rm cm^{-2}$, and keeping the normalisation and the spectral index as free parameters.

\subsection{\emph{Swift}-UVOT}
In addition to the X-ray data, Swift also provides an Optical/UV data via  \emph{Swift}-UVOT telescope \citep{2005SSRv..120...95R}. It observes  in optical and UV with the filters v, b, u and w1, m2, and w2 \citep{2008MNRAS.383..627P, 2010MNRAS.406.1687B}. We obtained the  data of BL\,Lac from \emph{HEASARC} Archive and reduced it to the scientific product by using the  \emph{HEASoft} package (v6.26.1).  The \emph{uvotsource} task included in the \emph{HEASoft} package (v6.26.1)  were used to process the images. Multiple images in the filters were added by using the \emph{uvotimsum} tool. The source counts were
extracted  by using a  circular source region of radius 5$''$ centred at the source location and  background region from a nearby source free circular region of radius 10$''$.  The observed fluxes were de-reddened for Galactic extinction using $\rm E(B-V)=0.2821$ and $R_{V} = A_{V}/E(B-V)=3.1$ following \citet{2007ApJ...663..320F}. We further corrected the flux densities for the contribution of the host galaxy by using the method outlined in  \citet{2013MNRAS.436.1530R}.  This prescribed method recommends  flux density values of 2.89, 1.30,  0.36, 0.026, 0.020 and 0.017 mJy for the host galaxy in the v, b,  u, uvw1, uvm2 and uvw2 bands, respectively. Moreover, the method suggests to subtract $\sim 50\%$ of the total host flux density to the observed flux density.

\section{Results}\label{sec:results}
BL\,Lac  has shown number of flaring events over the last  years across the energy bands,  these events  have been  reported in number of astronomical telegrams   
\citep[some  recent  telegrams are][]{2021ATel14777....1B, 2021ATel14826....1B, 2022ATel15688....1L, 2022ATel15730....1P}.
Recently, during the renewed activity from BL\,Lac,  the  $\gamma$-ray flux $\sim \rm  5\times 10^{-6}\rm~ph\,cm^{-2}\rm s^{-1}$ had been reported from the source \citep{2022ATel15688....1L}.   BL\,Lac  has also shown several bright VHE gamma-ray flares \citep{2020ATel14032....1B, 2021ATel14826....1B, 2021ATel14783....1C}. 
Motivated by the renewed flaring activity,  significant variability  and the availability of multiwavelength observations of BL\,Lac, we carried a detailed multiwavelength study of BL\,Lac by using the \emph{Fermi}-LAT and \emph{Swift}-XRT/UVOT observations. The aim is to understand the temporal and spectral characteristics of source. 

\subsection{Temporal Study}
\begin{figure*}
		\begin{center}
        \includegraphics[scale=0.7]{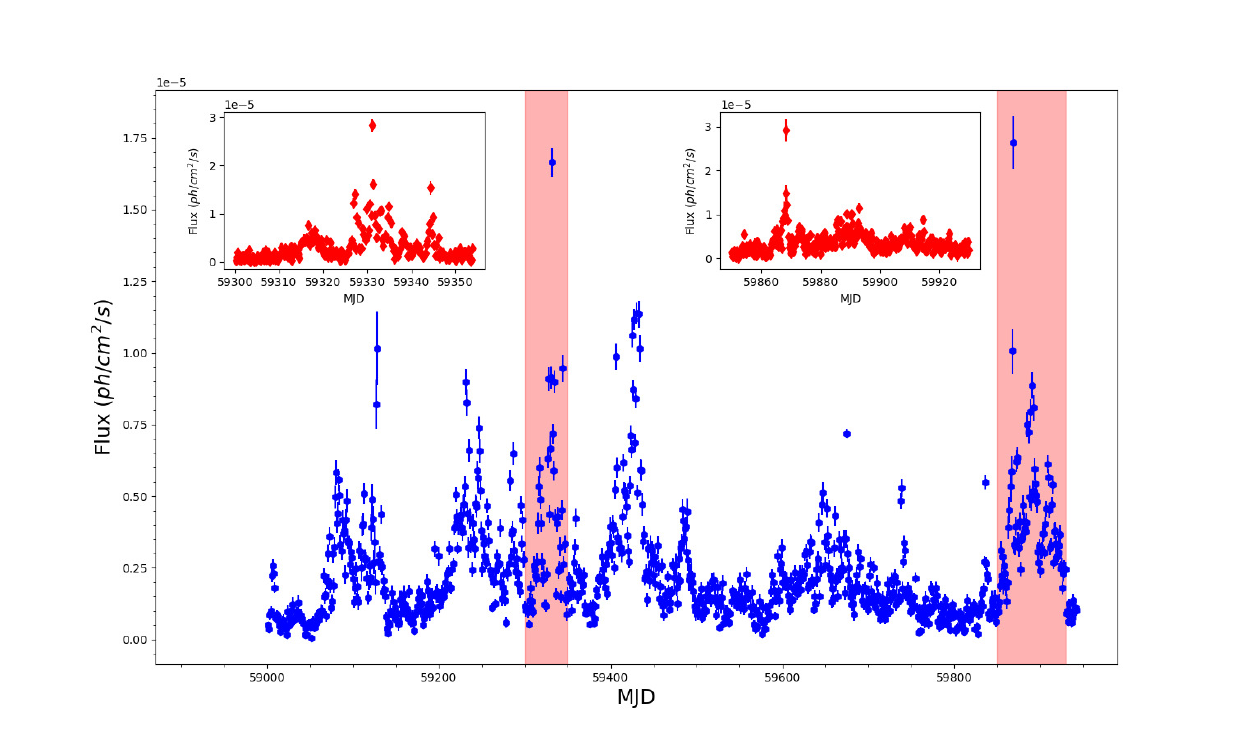}
        \caption{ One-day binned integrated \emph{Fermi}-LAT $\gamma$- ray light curve of BL\,Lac obtained  during the period MJD 59000--59943. The shaded strips represent the time periods for which the 6 hour binned $\gamma$-ray light curves are obtained. The inset plots display the 6 hour binned $\gamma$-ray light curves.}   
        \label{fig:gamma_lc}
		\end{center}        
        \end{figure*}

\begin{figure*}
		\begin{center}
        \includegraphics[scale=0.7]{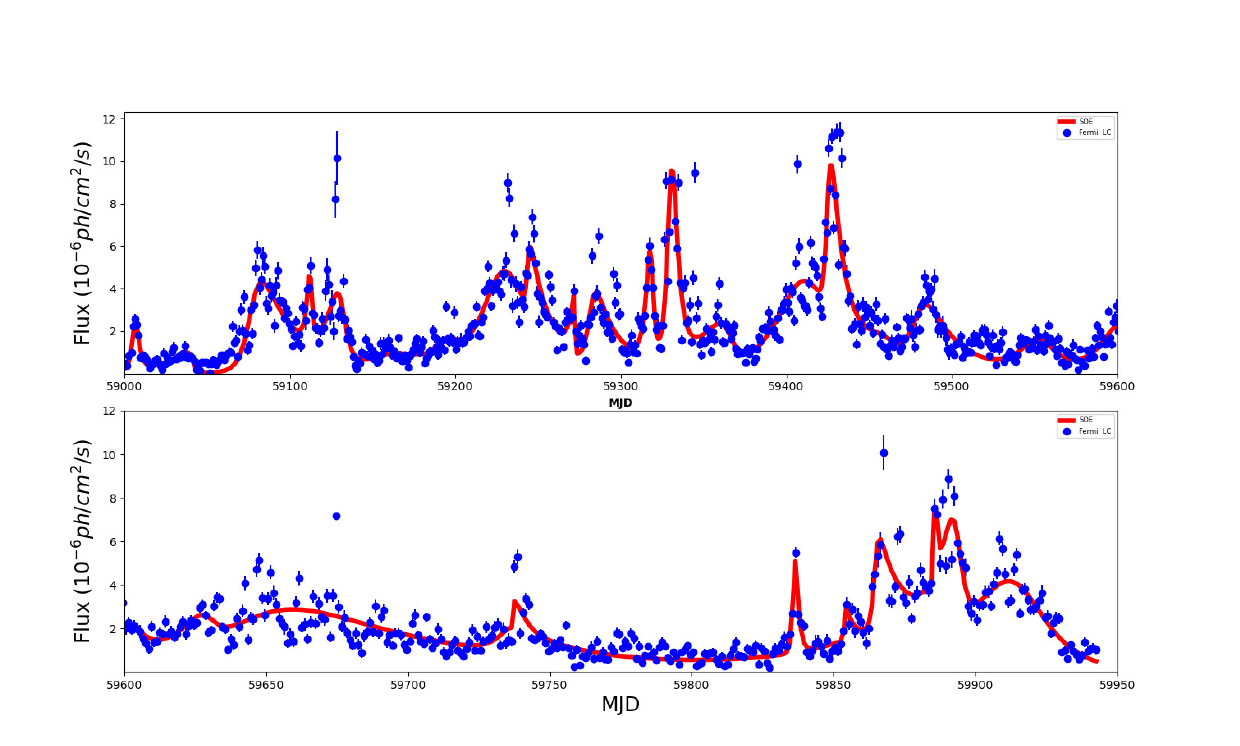}
        \caption{Daily binned $\gamma$-ray light curve of BL\,Lac fitted with the SOE function defined in Equation \ref{eq:rise_fall} .}       
        \label{fig:daily_lc_soe}
		\end{center}        
        \end{figure*}

We first analysed the \emph{Fermi}-LAT data of BL\,Lac acquired during the period MJD 59000--59943.  We obtained a daily binned $\gamma$-ray light curve  by integrating the photons over the energy range 100 MeV-500 GeV.   Specifically, the daily binned differential flux of  BL\,Lac was modelled  by power-law (PL) model
\begin{equation} \label{eq:fermi_pl}
\frac{dN}{dE}=N_0\left(\frac{E}{E_0}\right)^{-\Gamma}
\end{equation}
where $N_0$ is  the prefactor, $\Gamma$ is  the spectral index and $E_0$ is the scale energy. During  analysis, $N_0$ and $\Gamma$ were kept free in the fitting process, while $\rm E_0$ is chosen as $\rm 870\,MeV$.  We consider the source detection if TS obtained from the maximum-likelihood analysis exceeds 9, which corresponds to approximately a $\rm 3\,sigma$ detection level  \citep{2012ApJS..199...31N}. In all the time bins, we have ensured the convergence of the likelihood fit. The  resultant daily binned $\gamma$-ray light curve acquired is shown in Figure \ref{fig:gamma_lc}, all points displayed on the light curve exhibit a $3\,sigma$ detection.
As shown in Figure \ref{fig:gamma_lc}, the source started a major $\gamma$-ray activity around MJD\,59800 and  continued  high activity for more than  two and half years.
 During the active period,  a maximum daily averaged $\gamma$-ray flux of $(1.74\pm 0.09)\times10^{-5}\rm ph\,cm^{-2}\,s^{-1}$ is observed on  MJD 59868.5,  with a corresponding spectral index of $1.97\pm 0.04$ and  a TS value of 8008.  This is the highest ever $\gamma$-ray flux detected from BL\,Lac and is factor of $\sim 300$ larger than the base flux $F_0\sim 4.96\times 10^{-8} \rm ph\,cm^{-2}\,s^{-1}$, which is the minimum flux detected during this period.  The daily binned $\gamma$-ray light curve shows that the two maximum flux values are outliers. To check the accuracy of these flux values, we  perform further analysis by obtaining the 6 hour binned $\gamma$-ray light curves around these flux values, which are depicted by the shaded strips in the Figure \ref{fig:gamma_lc}.  We noted that the 6 hour binned light curves (shown in the inset plots) exhibited a similar pattern, with peak flux values of $(2.83\pm 0.12)\times10^{-5}\rm ph\,cm^{-2}\,s^{-1}$ and $(2.92\pm 0.25)\times 10^{-5}\rm ph\,cm^{-2}\,s^{-1}$ observed on MJD 59331.12 and MJD 59868.37, respectively.
 Moreover, the $\gamma$-ray light curve shows large flux variations with many low and high flaring  components.  We calculated the rise and decay times of these components by using the sum of exponential (SOE) function

\begin{equation}\label{eq:rise_fall}
F(t)=F_b+\Sigma F_{i}(t)\quad,
\end{equation}
where $F_b$ is the base line flux and
\begin{equation}
F_{i}(t)=\frac{2F_{p,i}}{\exp\left(\frac{t_{p,i}-t}{\tau_{r,i}}\right)+\exp\left(\frac{t-t_{p,i}}{\tau_{d,i}}\right)} \quad, \nonumber
\end{equation}
Here $F_{p,i}$ is peak flare amplitude at time $t_{p,i}$; $\tau_{r,i}$ and $\tau_{d,i}$ are  rise and decay times of the respective flare component. The fitted SOE profile along with the daily binned $\gamma$-ray light curve points are shown of Figure \ref{fig:daily_lc_soe}.
 We used more than 25 exponentials in the SOE function,  which resulted in $\chi^2$/dof of 8754/816.  We noted that adding more exponential in the SOE does not improve the fit statistics significantly. The best fit parameters of the components for which the peak flux is greater than $5\times10^{-6} ph\,cm^{-2}s^{-1}$ are given in Table \ref{table:multi-comp}. Using  rise/decay time scales,
we determined the profile shapes of  these components  by calculating the parameter,  $\zeta=\frac{\tau_d-\tau_r}{\tau_d+\tau_r}$,  such that the component  is symmetric if  $|\zeta|<0.3$,  moderately asymmetric if $0.3 <|\zeta| < 0.7$, and  asymmetric if   $0.7 < |\zeta| < 1$.  We  noted that five components are moderately asymmetric, one component is asymmetric and six components are symmetric.

  \begin{table*}
  \caption{Rise and fall time of  dominant  components of the light curve obtained by fitting the SOE  (Equation \ref{eq:rise_fall}) . Col. 1: peak time (MJD); 2: peak flux (in units $\rm 10^{-6} ph\,cm^{-2}{s^{-1}}$);   3 and 4:  rise time and decay time of the components (in days); 5: asymmetry parameter.}
\begin{adjustbox}{width=0.5\textwidth,center=\textwidth}
\begin{tabular}{lcccr}
\bottomrule
 $\rm t_p$ & $\rm F_p$ &$\rm t_{r}$ & $\rm t_{d}$ & |$\zeta$| \\
\bottomrule

59077.81$\pm$0.05    &  6.72$\pm$0.30   & 4.09$\pm$0.15 &   19.37$\pm$1.86 & 0.65 \\
59112.33$\pm$0.05    &   5.74$\pm$0.68   & 1.78$\pm$0.47 &   0.98$\pm$0.30 & 0.28 \\
59131.02$\pm$0.98   &    5.52$\pm$0.68  &   8.08$\pm$1.67 &    2.73$\pm$0.56 & 0.49 \\ 
59227.13$\pm$3.84   &   9.12$\pm$1.27   & 9.94$\pm$3.13 &     15.52$\pm$5.06  & 0.22  \\ 
59318.26$\pm$0.33   &   9.33$\pm$0.56   &  2.69$\pm$0.34 &  1.37$\pm$0.21  & 0.33  \\
59330.25$\pm$0.25   & 16.92$\pm$0.49  &    2.06$\pm$0.15 & 3.45$\pm$0.25  & 0.25  \\
59409.52$\pm$3.48  &  8.50$\pm$0.60    & 13.73$\pm$1.72 & 15.97$\pm$6.74  & 0.07  \\
59425.20$\pm$0.28  & 10.84$\pm$0.80  &  1.33$\pm$0.18   &  7.32$\pm$1.22  & 0.69   \\
59836.21$\pm$0.22  & 8.07$\pm$0.53    &    0.61$\pm$0.13    & 1.09$\pm$0.17  & 0.28 \\
59864.62$\pm$0.45  &   5.63$\pm$0.94  &  0.77$\pm$0.24 &   6.66$\pm$3.23  & 0.79 \\
59893.52$\pm$0.90  &   8.14$\pm$1.18  & 6.56$\pm$3.91 &   1.75$\pm$0.42  & 0.58 \\
59913.77$\pm$2.44  &  8.01$\pm$0.41  &  13.83$\pm$5.57 &  9.87$\pm$1.07  & 0.17 \\
\bottomrule
\end{tabular}
\end{adjustbox}
\label{table:multi-comp}
\end{table*}

\begin{figure*}
		\begin{center}
        \includegraphics[scale=0.9]{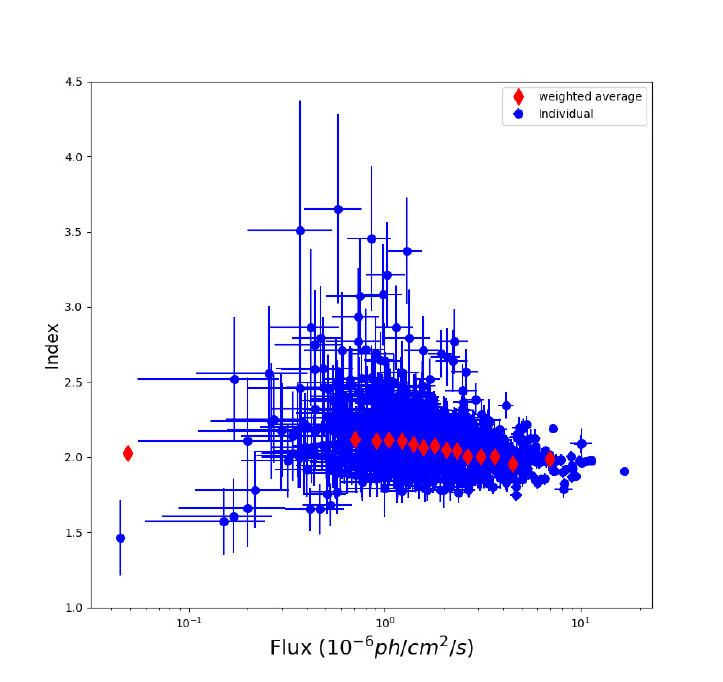}
        \caption{Daily binned $\gamma$-ray index  plotted as function of flux. Red diamond points represent the weighted average index values as function of weighted average flux values.}       
        \label{fig:index_flux}
		\end{center}        
        \end{figure*}

  The daily binned $\gamma$-ray light curve is obtained by fitting the  integrated spectrum with PL model.  In Figure \ref{fig:index_flux}, we plotted  the  index  as function of  daily binned $\gamma$-ray flux, the blue solid circles represent the individual flux-index points. We also sorted the flux and index values in the increasing order of flux and then average them over the bin of 13 points. In the Figure \ref{fig:index_flux}, red diamond points represent the weighted average index values as function of weighted average flux values. The Figure \ref{fig:index_flux} suggest  that  source exhibits a mild {\it harder when brighter} trend, a usual trend  observed in blazars \citep[e.g][]{2016ApJ...830..162B, 2019MNRAS.484.3168S}. We used the Spearman rank correlation method to calculate  the correlation coefficient and null-hypothesis probability between  index and flux values. The returned values of the correlation coefficient  $-0.37$ and null-hypothesis probability $4.52\times 10^{-31}$ further confirms a mild anti-correlation between the two quantities.
 
In order to obtain an estimate of shortest flux doubling  time scale, we scanned the daily binned $\gamma$-ray light curve with the equation
\begin{equation}
F(t)=F(t_0) 2^{\frac{t-t_0}{\tau}}\quad ,
\end{equation}
where $\rm F(t_0)$ and $\rm F(t)$ are the values of  flux at time $t_0$ and t respectively and $\tau$ is characteristic doubling time scale.
 On applying the condition that the significance of difference in flux at $\rm t$ and $\rm t_0$ is $\rm \geq 3\,sigma$ \citep{2011A&A...530A..77F}, the daily binned $\gamma$-ray light curve resulted in shortest flux doubling timescale,  $\rm t_{var}=0.40$ d  during the time interval  59022-59023.

\subsection{Fourier spectral analysis}

\begin{figure*}
		\begin{center}
        \includegraphics[scale=0.9]{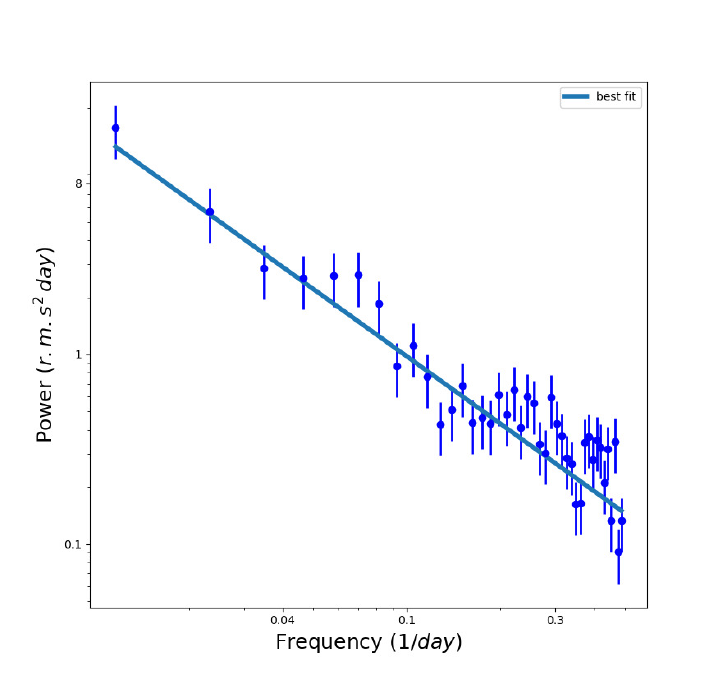}
        \caption{Power spectrum of one day binned $\gamma$-ray light curve. The fractional rms-squared normalization, also known as rms normalization is used in the plot.}
        \label{fig:power_spec}
		\end{center}        
        \end{figure*}

 Fermi-LAT provides longterm data of BL\,Lac object covering several years of observation.  As shown in Figure \ref{fig:gamma_lc}, the $\gamma$-ray light curve of BL\,Lac  shows unpredictable aperiodic variability with flux values changing  by several order in magnitude. Such random variability is often result of a stochastic processes, instead of  deterministic process. We used PSD to characterise  the variability.  PSD is one of the common tools used to examine the light curve variability,  it gives an estimate of  variability power as a function of temporal frequency.  In order  to obtain an insight into  the physical process which causes  large variability in  the  $\gamma$-ray light curve,  we obtained  PSD  of  source by splitting  daily binned $\gamma$-ray light curve into equal segments, and calculated  periodogram in each segment. We finally averaged them into  final periodogram. The averaged power spectrum is normalised by using the fractional r.m.s normalisation  \citep[see][]{1990A&A...227L..33B, 1992ApJ...391L..21M}.  
 The power spectrum is fitted with a  PL function,  $\rm P(f)=N\left(\frac{f}{f_0}\right)^{-\alpha_p} $ where N and $\rm \alpha_p$ are normalisation and spectral index respectively,  and $\rm f_0$ is the scale factor chosen as 0.01. The PSD points  along with best fit PL function ($\rm \chi^2_{red}=1.20$) are shown in  Figure \ref{fig:power_spec} and  best fit values of N and $\rm \alpha_p$  are obtained as $14.94\pm 2.76$ and  $1.18\pm 0.06$, respectively. The index $\sim 1$ of  power spectra  suggest for flicker-noise type process.  Similar results have also been  reported by \citep{2014ApJ...786..143S, 2020ApJ...891..120B}  in the  $\gamma$-ray light curves of few other blazars. 
The flicker noise  has the property of maintaining  shape over several orders of frequencies up to arbitrarily low values,  and thus observation of such features indicates long-memory process is at work.
 
\subsection{R.M.S flux relation and flux distribution study}
\begin{figure*}
		\begin{center}
        \includegraphics[scale=0.9]{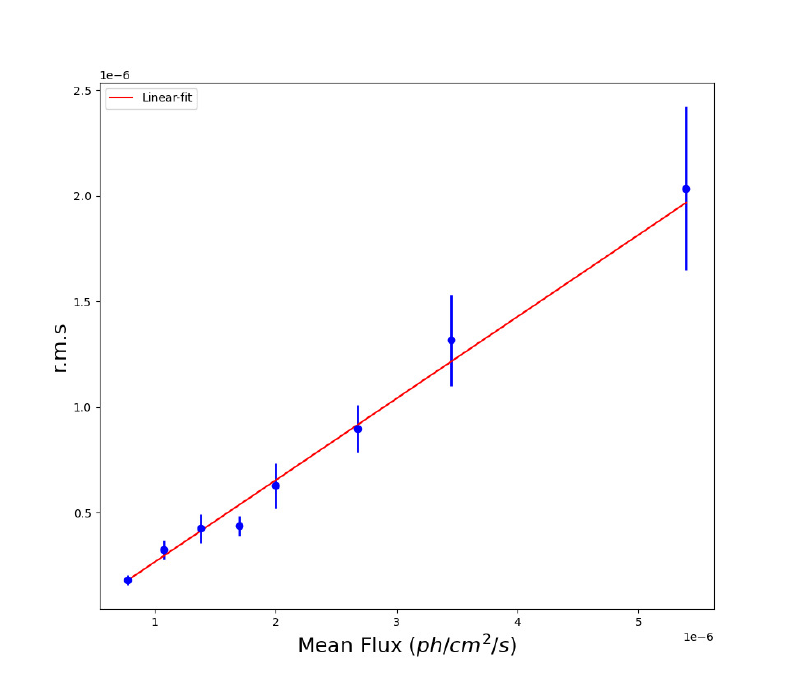}
        \caption{The variation of average absolute r.m.s values  as a  function of one day mean $\gamma$-ray flux (in units of $\rm ph\,cm^{-2}\,s^{-1}$).  Solid line represent the best linear fit. }       
        \label{fig:rms_flux}
		\end{center}        
        \end{figure*}

We  also characterised  the variability of  $\gamma$-ray light curve of BL\,Lac   by obtaining a correlation between the flux and r.m.s, commonly known as r.m.s--flux relation. Such relationships have increased our understanding of the variability in the astrophysical systems \citep[see e.g.][]{2001MNRAS.323L..26U, 2003MNRAS.345.1271V}.   In our work, we divide one day binned $\gamma$-ray light curve  into 80 segments of equal length such that each segment contains 11 data points. In each segment the mean flux and r.m.s values were calculated.  Next, the individual r.m.s values were sorted based on their corresponding mean flux and then binned into groups containing 8 values per bin. The resulting average r.m.s value for each bin is plotted as a function of the mean flux of that bin in Figure \ref{fig:rms_flux}, the plot includes a solid line that represents the best linear fit. The correlation plot  suggest  individual variance  track the source flux and  the variability responds to flux change on all measured timescales. 
 This trend has been noted in other astrophysical systems like Seyfert galaxies \citep{2001MNRAS.323L..26U, 2003MNRAS.345.1271V}. 
 
\subsection{Flux distribution}
\begin{figure*}
		\begin{center}
        \includegraphics[scale=0.8]{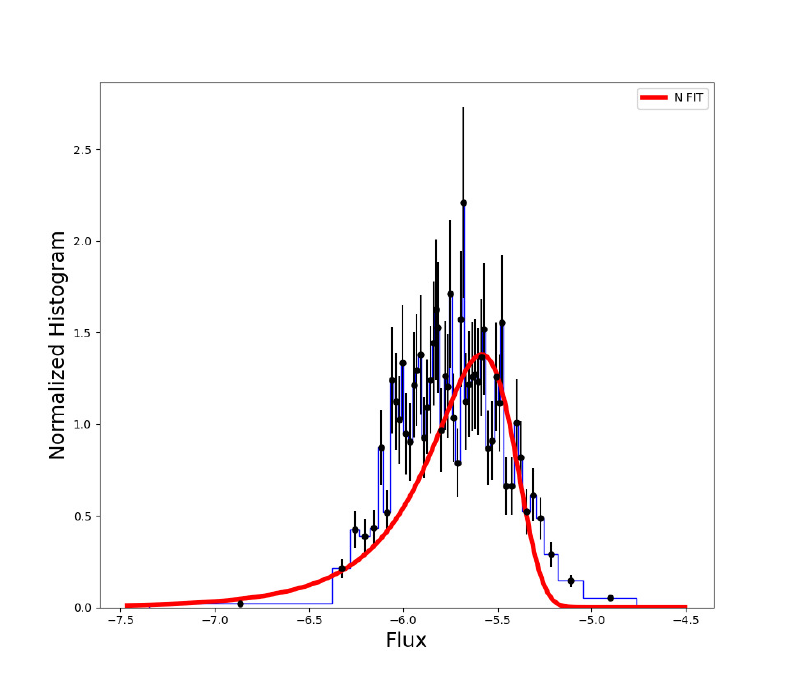}
        \includegraphics[scale=0.8]{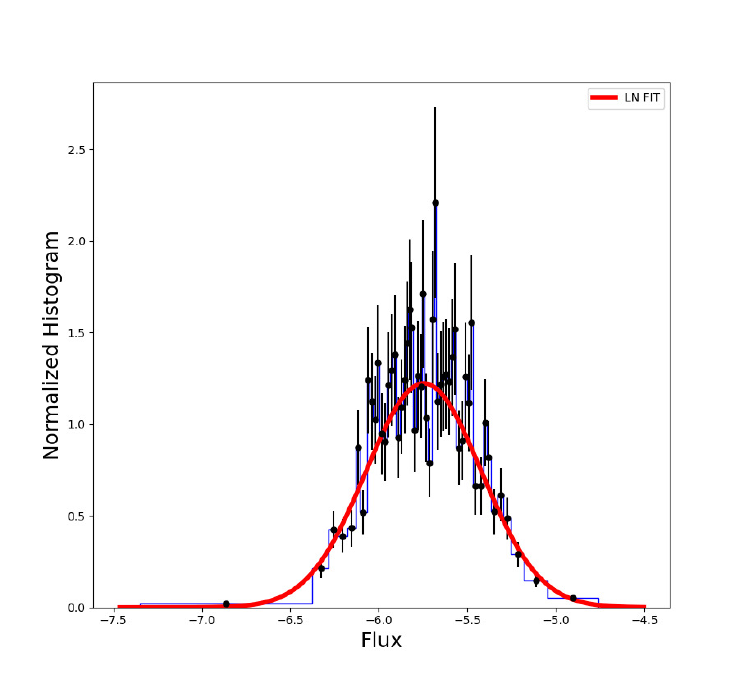}
        \caption{Left panel:  Normalised histogram of BL\,LAC fitted with the  Gaussian PDF. Right panel: Normalized histogram of BL\,LAC fitted with the   lognormal  PDF.  The flux are in units of $\rm ph\,cm^{-2}\,s^{-1}$.}       
        \label{hist_fit}
		\end{center}        
        \end{figure*}

Flux distribution study of the light curves of astrophysical systems  is another  tool used to probe the nature of underlying  physical processes responsible for the variability.  For example observation of normal flux distribution would indicate an additive processes, while a lognormal distribution would imply a multiplicative processes.
 Mostly the distribution of  observed flux of compact black hole systems follow a lognormal distribution.  
The linear r.m.s-flux relation  suggests that the flux distribution of the $\gamma$-ray light curve is lognormal. To investigate this,  we  characterise the $\gamma$-ray flux distribution of BL\,Lac  by performing the skewness, AD and histogram fitting test. The skewness and AD test rejects the normality of  flux distribution. 
Skewness value of the  flux distribution is obtained as $2.55\pm 0.16$,  and  the AD test yielded a  statistic value of 44.13, which is much larger than the  critical value (CV) of 0.78  defined at a 5\% significance level.  However, the AD test suggests the lognormality of the flux distribution.  The statistic value of 0.75 for the log of the flux distribution is smaller than $CV$ of  $ 0.78$  at 5\% significance level, which suggest  the null hypothesis that the flux distribution is  log-normal can not be rejected.
We further checked the PDF of flux distribution by constructing  the  normalised histogram  of logarithm of flux. The histograms were constructed such that each bin contains equal number of points while varying the bin width.  The normalised histogram points are plotted in Figure \ref{hist_fit}. The resulting histogram in log-scale is fitted by 

\begin{equation}\label{eq:ln}
L(x)=\frac{1}{\sqrt{2\pi}\sigma_l}e^{{-(x-\mu_l)^2/2\sigma_l^2}}
\end{equation}

and 

\begin{equation}\label{eq:nor}
G(x)=\frac{10^x \log(10)}{\sqrt{2\pi}\sigma_g}e^{-(10^x-\mu_g)^2/2\sigma_g^2}
\end{equation}

where $\mu_l$ and $\sigma_l$ are the mean and standard deviation of logarithm flux distribution,   $\mu_g$ and $\sigma_g$ are the mean and standard deviation of flux distribution.  Equation \ref{eq:ln} results in lognormal fit,  while Equation \ref{eq:nor} results in normal fit. The normalised histogram and best lognormal/normal fit to the histogram are shown in the left and right panel of Figure \ref{hist_fit}. 
The reduced-$\chi^2$ values obtained from fitting the flux distribution with a normal and a lognormal PDFs were 3.33 and 1.01, respectively. This result further suggests that the flux distribution is more accurately described by a lognormal distribution.
Observation of lognormal distribution in the $\gamma$-ray light curve suggest the underlying physical processes should be multiplicative in nature. It also suggest for two flux states in the source.

\subsection{Multiwavelength Light Curve}
\begin{figure*}
		\begin{center}
        \includegraphics[scale=0.8]{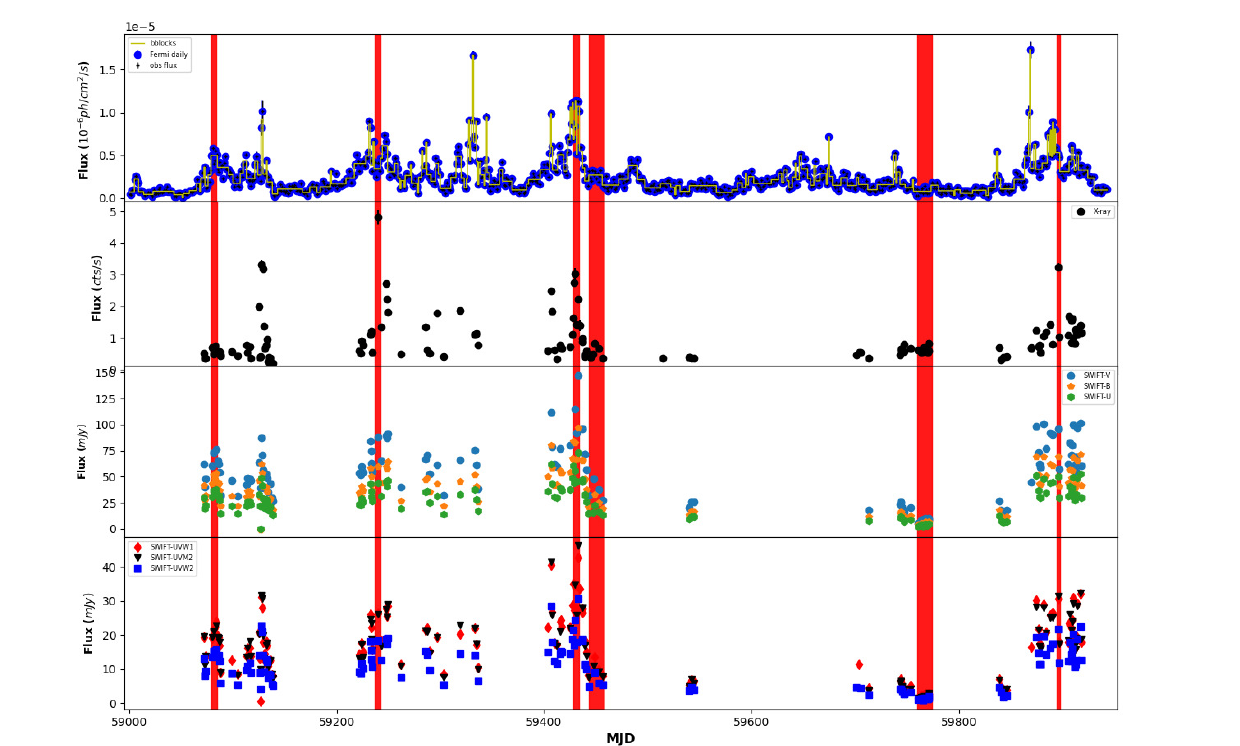}
        \caption{Multiwavelength light curves (MLC) of  BL\,Lac obtained by using \emph{Fermi}-LAT and \emph{Swift} XRT and UVOT observations. The observations spanned a  period from MJD 59000 to 59943. Top panel is daily binned $\gamma$-ray light curve, second, third and fourth panel are the X-ray, UV and optical light curves. The colored vertical stripes indicate the regions where broadband spectral modeling is performed.}       
        \label{fig:multiplot_lc}
		\end{center}        
        \end{figure*}

\begin{figure*}
		\begin{center}
        \includegraphics[scale=0.9]{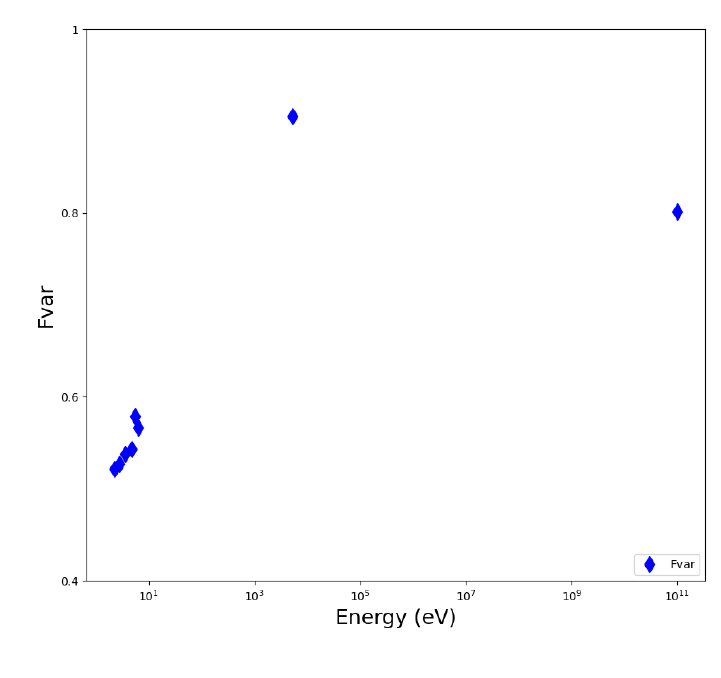}
        \caption{The fractional variability amplitude ($\rm F_{var}$) of BL\,Lac plotted as function of Energy }       
        \label{fig:ene_fvar}
		\end{center}        
        \end{figure*}

In order to unveil the behavior of BL\,Lac at optical, UV and X-ray band,  \emph{Swift}-XRT/UVOT had carried a total of 100 observations of  the source during the time period MJD 59000--59943. The X-ray and optical/UV data corresponding to these observation were analysed by following the analysis procedure described in section \ref{sec:data_ana}. The acquired  X-ray and optical/UV light curves of BL\,Lac  are shown in the second and bottom panels, respectively of the  multiwavelength light curve plot (see Figure \ref{fig:multiplot_lc}). Each point in the X-ray and optical/UV light curves  corresponds to individual observations.
The top panel shows the  daily binned $\gamma$-ray light curve with flux points obtained by integrating over the energy range 0.1--500 GeV.
The multiplot shows a simultaneous  flux variations in different energy bands.  We quantified  variability in each energy band by calculating  fractional variability amplitude using the equation \citep{2003MNRAS.345.1271V}
\begin{equation}\label{eq:fvar}
F_{var}=\sqrt{\frac{S^2-\overline{\sigma_{err}^2}}{\overline{F}^2}}
\end{equation}
where $\rm S^2$ is variance  and  $\rm \overline{F}$ is mean of flux points in the light curve, and $\rm \overline{\sigma_{err}^2}$ is  mean of  square of the measurement errors. The uncertainty on $\rm F_{var}$ is calculated using the equation \citep{2003MNRAS.345.1271V}
\begin{equation}
F_{var,err}=\sqrt{\frac{1}{2N}\left(\frac{\overline{\sigma_{err}^2}}{F_{var}\overline{F}^2}\right)^2+\frac{1}{N}\frac{\overline{\sigma_{err}^2}}{\overline{F}^2}}
\end{equation}
where N is the number of  points in the light curve. The values of  $\rm F_{var}$  acquired in the considered energy bands are plotted against the energy in Figure \ref{fig:ene_fvar}. The plot shows that the variability amplitude increases with energy from optical to $\gamma$-ray band,  the variability amplitude is highest in the $\gamma$-ray band.  This trend is  
can be manifestation of cooling of relativistic electron such that higher energy electrons  responsible for the $\gamma$-ray emission cool faster than the low energy electrons, which results in faster variability in the high energy emission.

\subsection{Broadband Spectral Analysis}
As shown in the  multi-wavelength plot (Figure \ref{fig:multiplot_lc}), the flux variation in different energy bands are correlated.  We employed the Z-transformed discrete correlation function  \citep[ZDCF,][]{1997ASSL..218..163A} to further explore the correlated variation between different  energy bands. The ZDCF values $+0.37, +0.62, +0.62, 0.63, 0.66, 0.59, 0.62$ and the time-lage values $-92.16_{-3.60}^{+90.12}$, $1.78_{-5.13}^{+3.22}, -2.22_{-2.54}^{+5.14}, -4.04_{-3.33}^{+5.26}$, $-1.02_{-2.81}^{+2.82}, 1.71_{-5.70}^{+2.52}, 2.98_{-5.29}^{+2.83} $ (the uncertainties are at 1 sigma confidence level) between the $\gamma$-ray light curve and the X-ray, B, U, V, W1, M2, W2 light curves, respectively suggest a  positive correlation between the emissions in different bands. Specifically, the acquired values indicate that the  $\gamma$-ray emission does not exhibits  a significant time lag with the X-ray and optical/UV emission.
 In order to identify the underlying particle energy distribution and  physical parameters (like magnetic field, size of emission region, bulk Lorentz factor etc) responsible for the simultaneous flux variations,  we  examined the  broadband spectral characteristics of BL\,Lac  by choosing different time intervals from the multiwavelength light curve. 
  Due to significant variability exhibited by BL\,Lac source on short timescales,  we utilized Bayesian analysis to divide the $\gamma$-ray light curve into segments, with each segment assumed to exhibit a steady  behavior in terms of the underlying physical parameters. This allows us to estimate the most probable values of the parameters for each segment. 
 We then selected the Bayesian segments for the broadband spectral analysis based on source activity and  simultaneous observations available in $\gamma$-ray, X-ray and UV/optical bands. We have shown these time segments  by the  vertical stripes in the multiplot (Figure \ref{fig:multiplot_lc})  and  identified them  as  
 S-1 MJD 59079-59084, S-2 MJD\,59237-592425, S-3 MJD\,59428--59434, S-4 MJD\,59894-59897 flux states, Quiescent States:  QS-1 MJD\,59443-59457 and QS-2  MJD\,59760-59774. 
  By comparing the results from different segments, one can identify changes in the physical  parameters that govern the flux variations.
 
\begin{figure*}
		\begin{center}
	\includegraphics[angle=270,width=.47\textwidth]{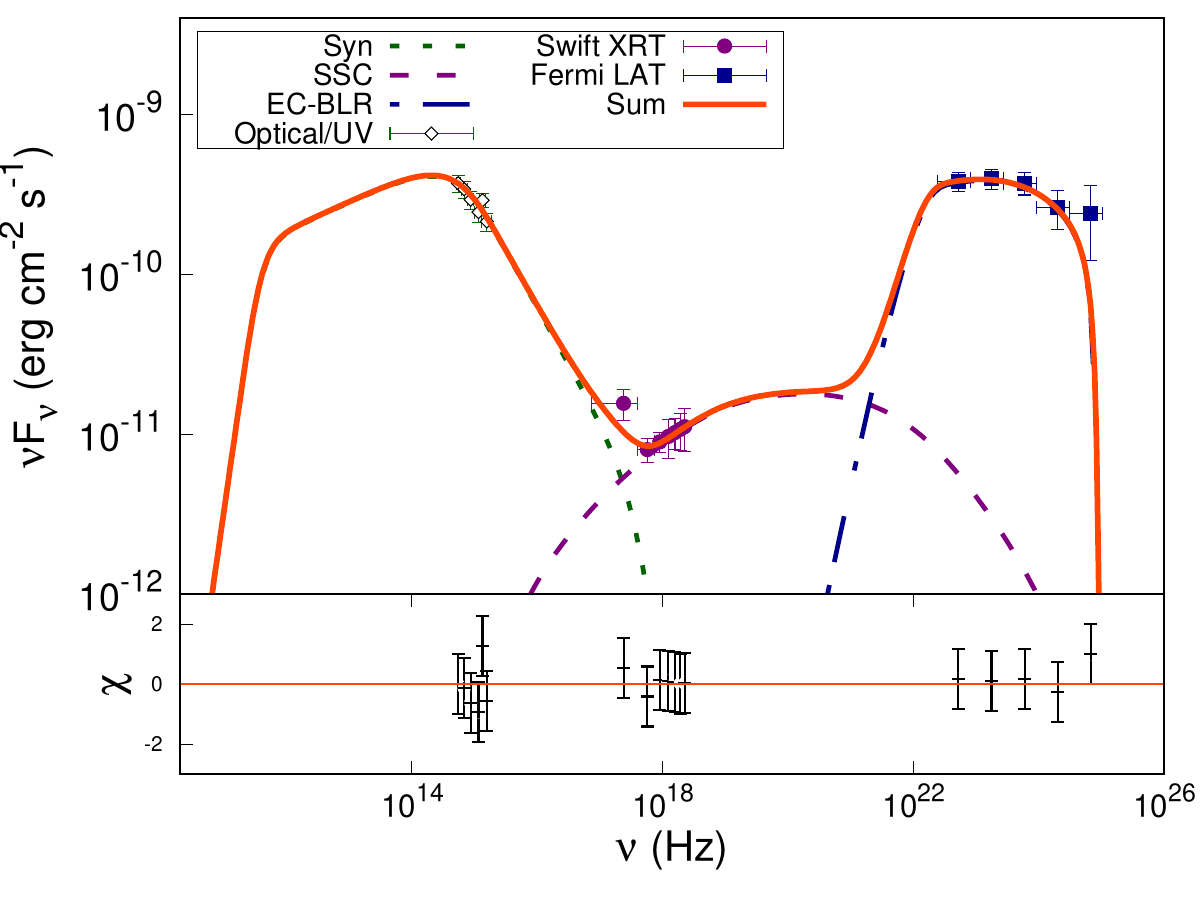}
        \includegraphics[angle=270,width=.47\textwidth]{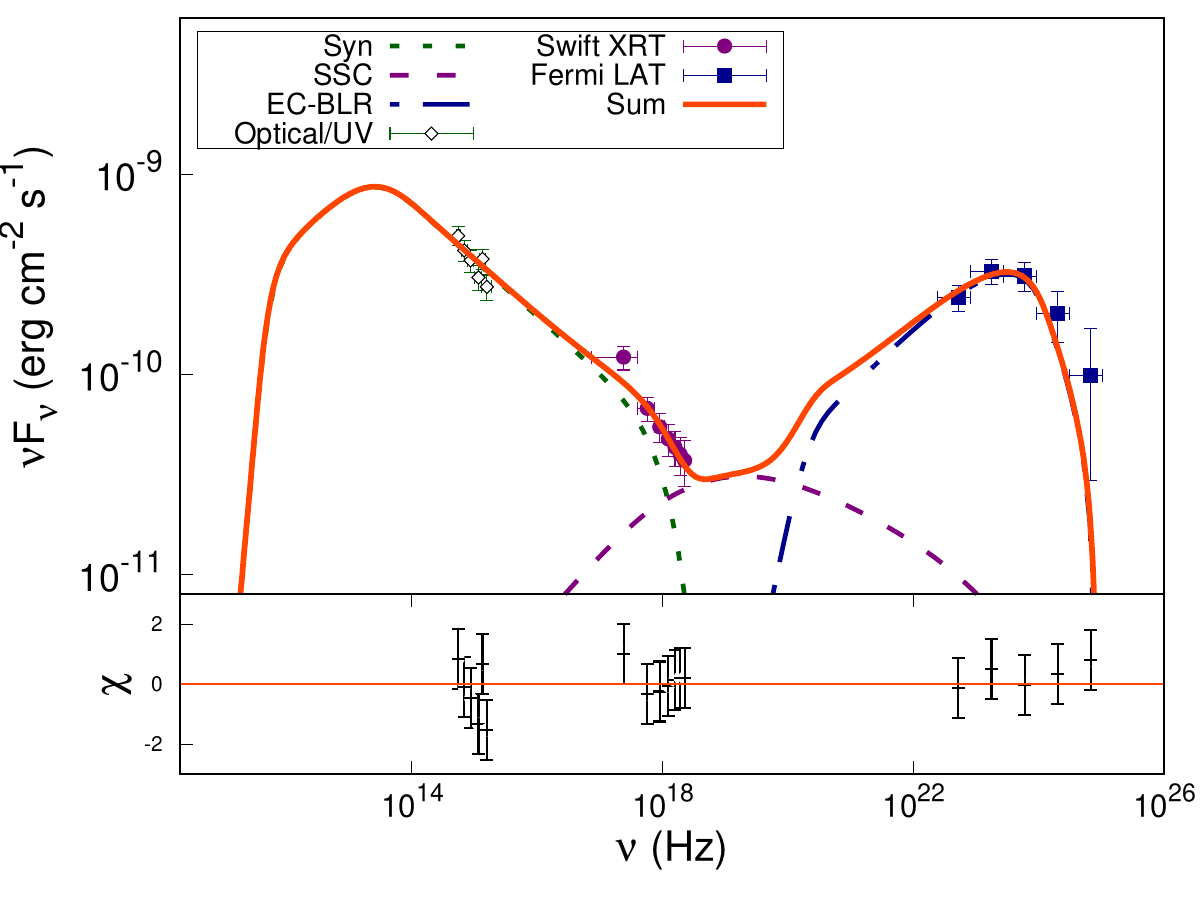}
	   \vspace{0.9cm}
        \caption{Broadband SED of BL\,Lac obtained during the flux state S-1 (left panel)and S-2 (right panel). The  flux points are represented  by open diamond (\emph{Swift}-UVOT), filled circles (\emph{Swift}-XRT), and filled squares (\emph{Fermi}-LAT). The solid red curve represents the combined best fit synchrotron, SSC and EC  spectrum.}
        \label{fig:s12}
		\end{center}        
\end{figure*}

\begin{figure*}
		\begin{center}
       \includegraphics[angle=270,width=.47\textwidth]{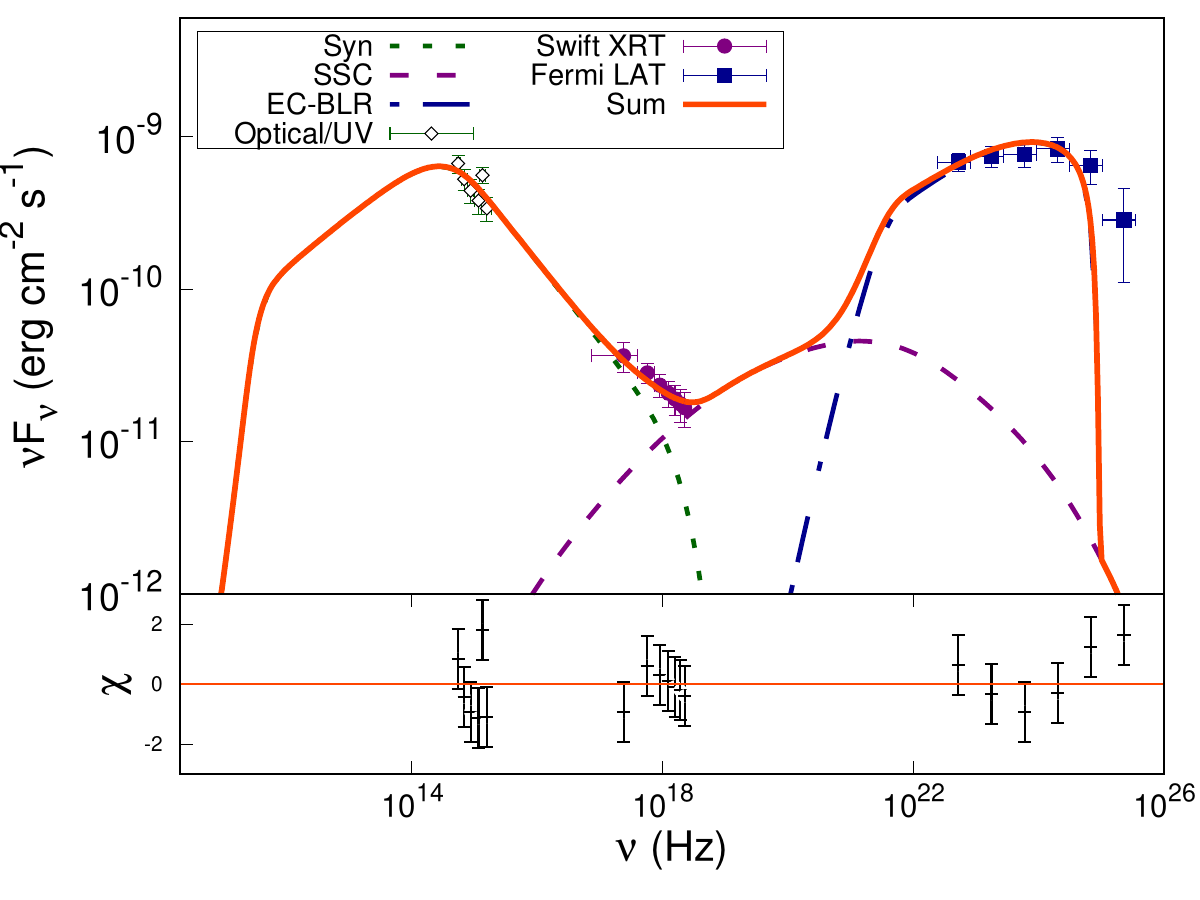}
        \includegraphics[angle=270,width=.47\textwidth]{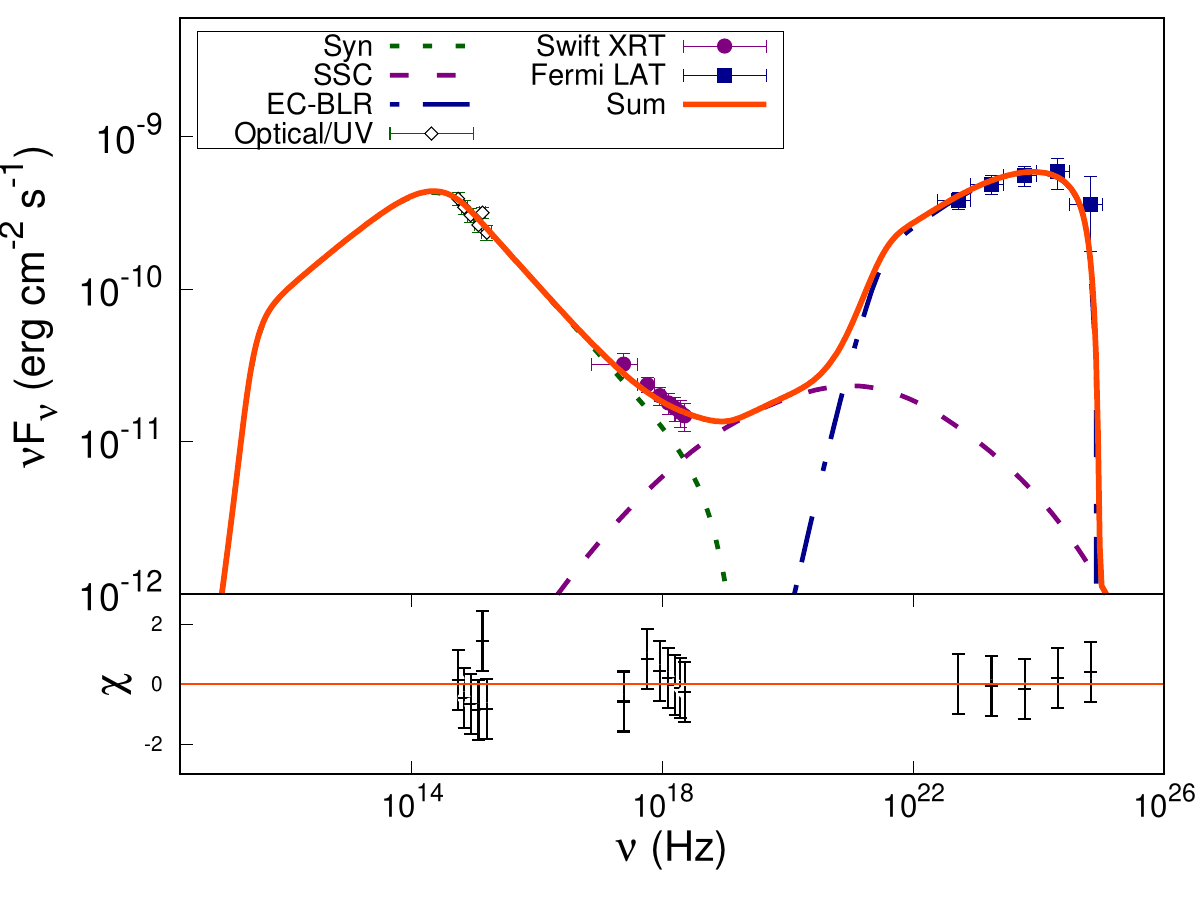}
	   \vspace{0.9cm}
        \caption{SED of BL\,Lac obtained during the flux state S-3 (left panel) and S-4 (right panel). The labelling are same as that of Figure \ref{fig:s12}.}
        \label{fig:s34}
		\end{center}        
\end{figure*}

\begin{figure*}
		\begin{center}
       \includegraphics[angle=270,width=.47\textwidth]{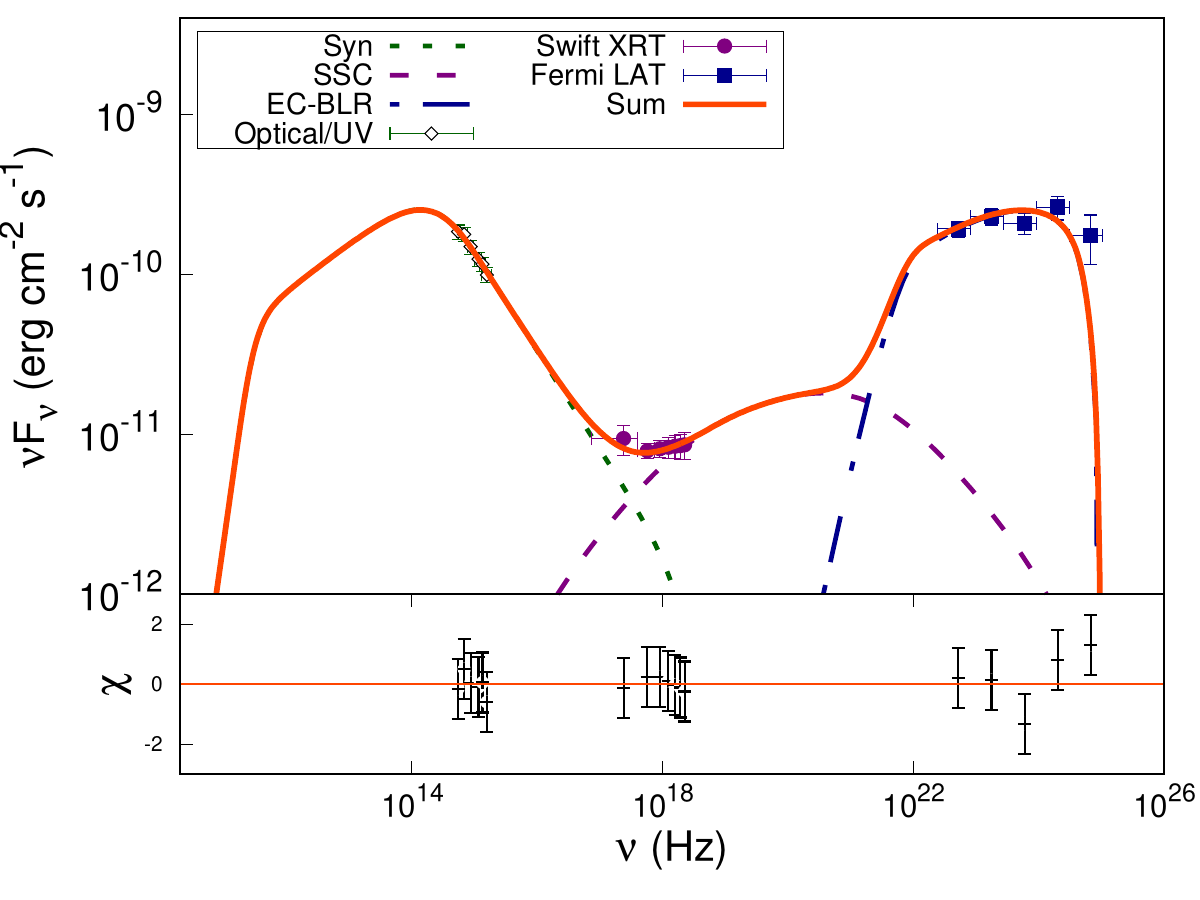}
       \includegraphics[angle=270,width=.47\textwidth]{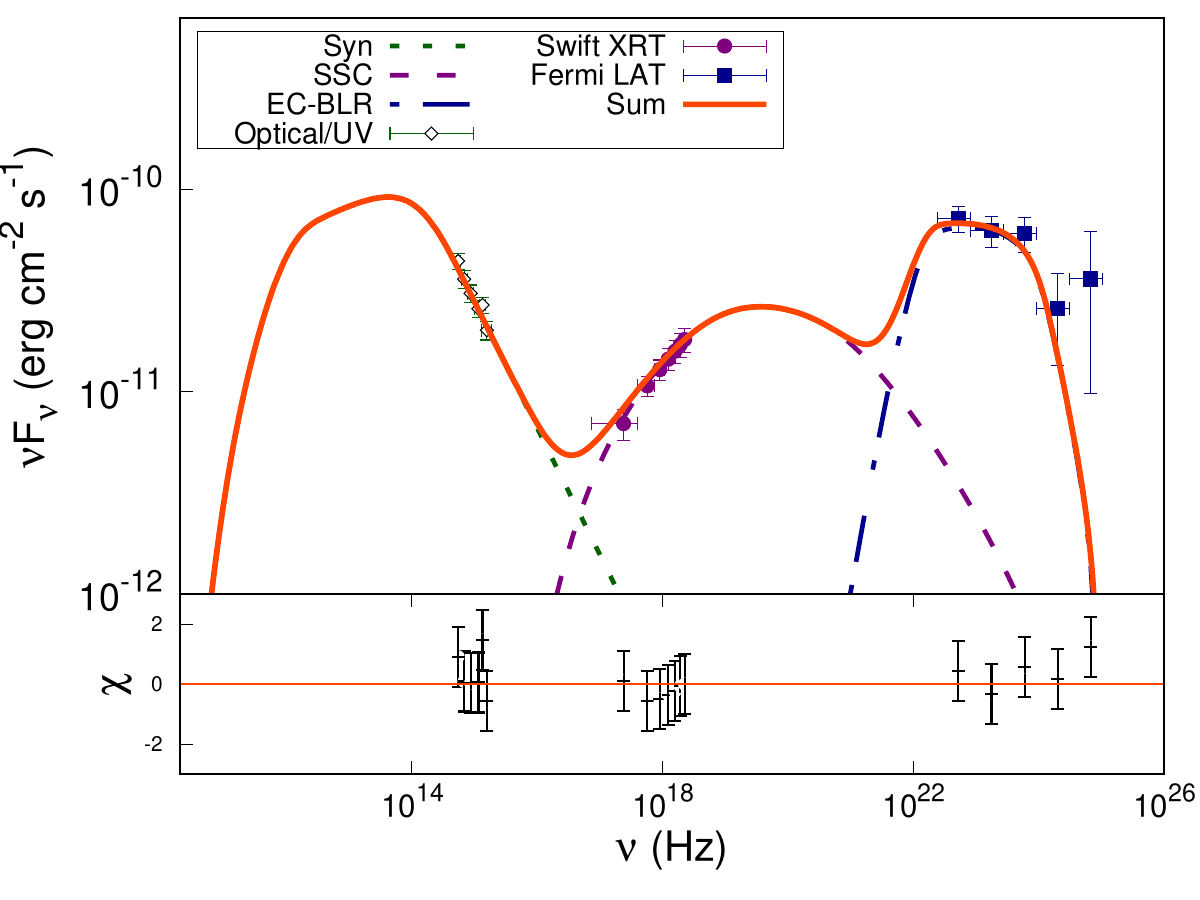}

	   \vspace{0.9cm}
        \caption{SED of BL\,Lac obtained during the QS-1 and QS-2 flux state. The labelling are same as that of Figure \ref{fig:s12}.} 
        \label{fig:q12}
		\end{center}        
\end{figure*}

We modeled the  $\gamma$-ray spectrum in the considered flux states with
LP model, defined in Equation \ref{eq:fermi_pl} and PL model,  $\rm \frac{dN}{dE}=N_0(E/E_0)^{-\Gamma}$, where $\rm N_0$ is the prefactor, $\Gamma$  denotes the spectral index and $E_0$ represents the scale energy, which is fixed at 856 MeV. 
The  fitting parameters obtained are  summarised in Table \ref{table:spec_fit_param}. 
We determined the statistical significance of the curvature in the $\gamma$-ray spectrum by using the relation $\rm TS_{curve}=2[\log\mathcal{L}(LP)-\log\mathcal{L}(PL)$] \citep{2012ApJS..199...31N}.
 As shown in Table \ref{table:spec_fit_param}, significant curvature ($\rm TS_{curve}>16$) is observed in S-2, S-3 and QS-1 states. The $\gamma$-ray  spectral points  for the broadband SED modelling are obtained by dividing  the total energy (1--500 GeV) into 8 energy bins equally spaced in log scale. For the S-2, S-3, and QS-1 states,  the source spectrum is fitted using a  LP model, while for the S-1, S-5, and QS-2 states, a PL model was employed.  During the spectral fit in each energy bin,  the spectral parameters of BL\,Lac were kept free, while the parameters of other  sources in the ROI were frozen to the best fit values acquired in the energy range 0.1--500 GeV. The X-ray spectrum in each flux state is obtained  by using an online automated products generator tool \citep{2009MNRAS.397.1177E}. We binned  the acquired X-ray spectrum by using the GRPPHA task,  such that their are 20 counts per bin. In case of \emph{Swift}-UVOT, the images of the observation IDs falling in particular flux state are combined using the UVOTIMSUM task,  and finally  flux values are obtained from the combined image. The broadband SED points obtained in the  S-1, S-2, S-3, S-4,  QS-1 and QS-2  are shown in Figures \ref{fig:s12}, \ref{fig:s34} and \ref{fig:q12}.

\begin{table*}
\caption{The parameters obtained by fitting the integrated $\gamma$-ray spectrum of  S-1, S-2, S-3, S-4,  QS-1 and QS-2 states of BL\,Lac with the PL and LP model. Col. 1: flux state; 2: time period of flux state; 3: fitted model; 4: integrated flux in units of $\rm~10^{-6}\,ph\,cm^{-2}\,s^{-1}$; 5: PL index or index defined at reference energy; 6: curvature parameter; 7: test statistics; 8: -log(likelihood); 9: significance of curvature.}
\begin{adjustbox}{width=1.0\textwidth,center=\textwidth}
\begin{tabular}{lcccccccr}
\bottomrule
State  & period & Model & $\rm F_{0.1-500}$ & $\Gamma$ or $\alpha$ & $\beta$ & TS & -$\log\mathcal{L}$ & $\rm TS_{curve}$ \\
\bottomrule
S-1 & 59079-59084 & PL & $2.38\pm 0.13$ & $2.05\pm 0.04$  & ---  & 1602  &  6322 & ---  \\
 &  & LP & $2.34\pm 0.06$ & $2.04\pm0.01$ & $0.07\pm0.01$ & 1663 & 6318 &  8\\
 
S-2 & 59237-59242 & PL & $1.73\pm0.11$ & $1.99\pm0.04$ & --- & 1269 & 5993  & ---\\
	& & LP & $1.59\pm 0.38$  & $1.94\pm 0.09$ & $0.13\pm0.04$ & 1305 & 5982 & 22\\
	
S-3 & 59428-59434 & PL &  $4.44\pm0.12$ & $1.98\pm0.02$ & --- & 7722 & 13109 & ---\\
      &  &  LP  &  $4.35\pm0.11$ & $1.96\pm0.02$  &  $0.03\pm0.01$  & 11940 & 13098  &  22\\
      
S-4 & 59894-59897 & PL &  $2.70\pm0.18$ & $1.93\pm0.04$ & --- & 1200 & 3990 & ---\\
      &  &  LP  &  $2.54\pm0.0.02$ & $1.88\pm0.005$  &  $0.07\pm0.002$  & 1229 &  3985 & 10 \\

QS-1 & 59443-59457 & PL & $1.33\pm0.06$ & $1.99\pm0.03$ & --- & 2428 &  11099 & ---\\
	& & LP & $1.25\pm0.05$  & $1.95\pm0.02$ & $0.05\pm0.01$ & 2407 & 11084 &  30\\
	
QS-2 & 59760-59774 & PL & $0.44\pm0.04$ & $2.17\pm0.07$ & --- & 398 &  9400 & ---\\
	& & LP & $0.43\pm0.02$  & $2.16\pm0.02$ & $0.03\pm0.01$ & 402 & 9398 & 4 \\

\bottomrule
\end{tabular}
\end{adjustbox}
\label{table:spec_fit_param}
\end{table*}

We consider one-zone leptonic  model in order to model  broadband SED  in the selected flux states.  In this model, we assume emission arises from  a spherical blob of radius $\emph{R}$  filled with  relativistic electron distribution, $\rm n(\gamma)$. The  blob  moves down the jet with  bulk Lorentz factor $\Gamma$ at a  small angle $\theta$ with respect to line of sight of observer. Relativistic motion at small angle with respect to observer amplifies the blazar emission and this amplification is determined by a beaming factor $\delta = 1/\Gamma(1-\beta\cos\theta)$, $\beta$ is the velocity (in units of c) of blob.  Further, we assume that the variability is governed by  light crossing time scales so that size of emission is obtained through expression $\rm R\sim \delta t_{var}/(1+z)$. 
 The relativistic electrons  in presence of  magnetic field, B and target photon field  emit radiations through synchrotron and IC processes. In our model, we assume seed photons for IC process are synchrotron photons from jet itself, so that emission is through SSC process. We expressed  the electron Lorentz factor $\gamma$ in terms of new variable  $\xi$ such that $\xi =\gamma\sqrt{\mathbb{C}}$, where $\rm \mathbb{C} =1.36\times 10^{-11}\delta B/(1+z)$.  Following \citet{1984RvMP...56..255B},  synchrotron flux at energy $\epsilon$ can be obtained using the equation
 
 \begin{equation}\label{eq:syn_flux}
 F_{syn}(\epsilon)=\frac{\delta^3(1+z)}{d_L^2} V  \mathbb{A}  \int_{\xi_{min}}^{\xi_{max}} f(\epsilon/\xi^2)n(\xi)d\xi,
 \end{equation}
 
 where, $\rm d_L$  is  luminosity distance, V is volume of  emission  region, $\rm \mathbb A = \frac{\sqrt{3}\pi e^3 B}{16m_e c^2 \sqrt{\mathbb{C}}}$,  $\xi_{min}$ and $\xi_{max}$ correspond to the minimum and maximum energy of electron,  and f(x) is the synchrotron emmisivity function \citep{1986rpa..book.....R}. The
 SSC flux received by the observer at energy $\epsilon$  can be obtained using the equation 
 \begin{equation}\label{eq:ssc_flux}
  \begin{split}
 F_{ssc}(\epsilon) =\frac{\delta^3(1+z)}{d_L^2} V  \mathbb{B} \epsilon & \int_{\xi_{min}}^{\xi_{max}} \frac{1}{\xi^2}  \int_{x_1}^{x_2}   \frac{I_{syn}(\epsilon_i)}{\epsilon_i^2}  \\
&  f(\epsilon_i, \epsilon, \xi/\sqrt{\mathbb{C}}) d\epsilon_i   n(\xi)d\xi
 \end{split}
 \end{equation}
where, $\rm \epsilon_i$ is incident photon energy,   $\rm \mathbb{B} = \frac{3}{4}\sigma_T\sqrt{\mathbb{C}}$,  $\rm I_{syn}(\epsilon_i)$ is the synchrotron intensity,  $\rm x_1=\frac{\mathbb{C} \, \epsilon}{4\xi^2(1-\sqrt{\mathbb{C}} \,\epsilon/\xi m_ec^2)}$,  $\rm x_2=\frac{\epsilon}{(1-\sqrt{\mathbb{C}}\,\epsilon/\xi m_e c^2)}$ and

\begin{equation}
f(\epsilon_i, \epsilon, \xi)= 2q\log q+ (1+2q)(1-q)+\frac{\kappa^2q^2(1-q)}{2(1+\kappa q)} \nonumber
\end{equation}
here $\rm q=\frac{\mathbb{C}\epsilon}{4\xi^2\epsilon_i(1-\sqrt{\mathbb{C}}\epsilon/\xi m_ec^2)}$ and $\rm \kappa=\frac{4\xi\epsilon_i}{\sqrt{\mathbb{C}} m_e c^2}$.

  Similarly the EC flux recieved by the observer can be obtained using the equation 
 
 \begin{equation}\label{eq:ec_flux}
 \begin{split}
 F_{ec}(\epsilon) =\frac{\delta^3(1+z)}{d_L^2} V  \mathbb{D} \epsilon &  \int_{0}^{\infty}  d\epsilon_i^* \int_{\xi_{min}}^{\xi_{max}} d\xi \frac{N(\xi)}{\xi^2} \frac{U_{ph}^*}{\epsilon_i^*} \eta(\xi,\epsilon_s, \epsilon_i')  \\
 \end{split}
 \end{equation}
 where $\rm \mathbb{D}=\frac{3}{32\pi}c\beta\sigma_T\sqrt{\mathbb{C}}$,  $\rm \epsilon_i^*$ is the energy of target photons in the AGN frame, $\rm U_{ph}^*$ is target photon energy density and 
 
 \begin{equation}
 \eta(\xi,\epsilon_s, \epsilon_i') =y+\frac{1}{y}+\frac{\mathbb{C}\epsilon_s^2}{\xi^2\epsilon_i'^2y^2}-\frac{2\nu_s\sqrt{\mathbb {C}}}{\xi\epsilon_i'y}
 \end{equation}
 here $\rm y=1-\frac{\sqrt{\mathbb{C}} \epsilon_s}{\xi m_e c^2}$.
 
 We solved Equations \ref{eq:syn_flux},  \ref{eq:ssc_flux} and  \ref{eq:ec_flux} numerically and the resultant  numerical code is incorporated as a local convolution model in XSPEC in order to perform a statistical fitting of  broadband SEDs. The convolution code allows us to model the broadband spectrum for any particle energy distribution, $n(\xi)$.  In a single XPEC spectral fitting iteration,  the synchrotron and IC ( (SSC/EC) process are solved simultaneously and same set of parameters (e.g B, $\Gamma$, $\rm R$ etc) are utilized in these process in a single itteration.
 In our convolution code,  XSPEC `energy' variable is  interpreted as $\xi = \gamma \mathbb{C}$. Three cases of particle distribution viz.,  broken power law (BPL),   LP, and the  physical model namely energy dependent acceleration \citep[EDA; ][]{2021MNRAS.508.5921H} were considered in our analysis. We first examined the underlying particle distribution  by modelling broadband SED  of  S-1 flux state.  Our results show that the convolved  BPL electron distribution undergoing synchrotron and SSC loses provides better fit  ($\chi^2/dof \sim$ 4.62/14) to the broadband SED than the LP and EDA model with  $\chi^2/dof$ of 30.45/14 and 32.83/14 respectively. 
 This suggest that the underlying particle distribution responsible for  broadband emission in BL\,Lac is preferably a  BPL model.  Hence,  we used BPL form of electron distribution to fit the  broad band SED in the considered flux states.  A systematic of 10\% is added evenly over the entire data  to account for the additional  uncertainties in the model. Using the convolved SED model involving synchrotron and SSC processes with BPL electron distribution, the observed broadband spectrum is determined mainly by 10  parameters viz. $\rm \xi_b$, $\rm \xi_{min}$, $\rm \xi_{max}$,   p, q, $\Gamma$, B, R,  $\theta$  and norm N. 
 The code also allows us to fit the SED with jet power ($\rm P_{jet}$) as one of the parameter, however in this case,  N must be a fix parameter. 
 We carried fitting with p, q,  $\Gamma$,  and B as free parameters, while other parameter were freezed to typical values  required by the observed broadband spectrum. The reason for  freezing the parameters is limited information available in the Optical/UV, X-ray and $\gamma$-ray bands. Moreover, we used Tbabs model to account for the absorption in the X-ray spectrum.  We noted that synchrotron and SSC emission provided a  reasonable fit to all the flux states with  $\rm \chi^2/dof$ values as 4.62/14, 14.96/14, 12.05/15, 8.16/14, 12.06/14, and 6.18/14 for S-1, S-2, S-3, S-4, QS-1 and QS-2 states, respectively. However, the equipartition values obtained as 2252, 216, 16100, 9060, 2096, 3864 in the S-1, S-2, S-3, S-4, QS-1 and QS-2  flux states are much larger than unity. The results indicate additional process responsible for the high energy emission.  The BL Lac source is well-known for exhibiting emission line features originating from the BLR  \citep{1995ApJ...452L...5V, 1996MNRAS.281..737C}. These results indicate that the BLR region may play significant role for the high energy emission. Therefore, we  investigated the  broad band emission  from BL\,Lac by considering the EC emission alongside synchrotron and SSC emissions. We assumed the seed photons for EC scattering are BLR photons. Interestingly, the emission lines detected from the source mainly consists of  $\rm H_\alpha$ emission lines  \citep{1995ApJ...452L...5V, 1996MNRAS.281..737C}, hence
 for numerical stability, we approximate the BLR emission as a blackbody with a temperature 42000 K (equivalent to the temperature corresponding to Lyman-alpha line emission at $2.5 \times 10^{15}$ Hz). Using this model, the observed broad-band spectrum can now be reproduced by introducing two additional parameters viz. target photon temperature $T$ and  fraction of the photons which undergo Compton process $\rm f$. 
 The limited observation information and large number of model parameters introduces a degeneracy in  model parameters. The optimum set of parameters would be those for which  $\rm P_{jet}$ is minimum. 
To check for minimum $\rm P_{jet}$,  we varied  $\rm \xi_{min}$, which corresponds to variation in $\rm \gamma_{min}$ through the relation  $\rm \gamma_{min}=\xi_{min}/\sqrt{C}$.
The $\rm P_{jet}$ and the $\rm \chi^2/dof$ obtained for different values of $\rm \gamma_{min}$ are given in the Table \ref{table:jet_power}.  In all the flux states,   $\rm P_{jet}$ shows decreasing trend as $\rm \gamma_{min}$ increases, and the reduced-$\chi^2$ obtained is reasonable good up to $\rm \xi_{min}\sim 10^{-3}$.  Therefore, keeping the minimum jet power in mind,  the value of  $\xi_{min}$ is constrained between $10^{-3} -10^{-4}$ in all the flux states in the final SED fit. 
The resultant best-fit model SED along with observed points are shown in Figures \ref{fig:s12}, \ref{fig:s34}, and \ref{fig:q12}
 and corresponding best fit parameters are given in Table \ref{table:sed}. 

\begin{table*}
\caption{The table shows variation of the Jet power with the $\rm \gamma_{min}$ obtained by using the local convolution SED model. Col: 1. Flux state,  2. $\rm \xi_{min}$ parameter, 3. $\rm \gamma_{min}$, 4. Jet power and 5. $\rm \chi^2/dof$.}
\begin{adjustbox}{width=0.5\textwidth,center=\textwidth}
\begin{tabular}{lccccc}
\bottomrule
Flux state  & $\rm \xi_{min}$ & $\rm \gamma_{min}$  & $\rm P_{jet}$ &   $\rm \chi^2/dof$  \\
\bottomrule
	&  $10^{-6}$ &  0.07 & 48.78  & 15.78/14 \\
	&  $10^{-5}$ &  0.75  & 47.48  & 16.03/14 \\
S-1	& $10^{-4}$  & 6.08 & 46.51   & 5.33/14  \\
	&  $10^{-3}$  & 64.64  & 45.02  & 3.74/14 \\
	\vspace{1.5mm}
	&   $10^{-2}$  &  1004 &  44.43    & 28.04/14 \\
\hline	
	&  $10^{-6}$ &  0.05  & 49.18   & 7.16/14\\
	&  $10^{-5}$ &  0.50  & 47.79  & 7.15/14 \\
S-2	&  $10^{-4}$  & 5.19 & 46.44   & 7.37/14  \\
	&  $10^{-3}$  & 53 & 45.18   &  7.42/14 \\
	\vspace{1.5mm}
	&   $10^{-2}$ & 859  & 44.07  &25.12/14 \\	
\hline		
	&  $10^{-6}$  &  0.69 & 48.52  & 16.03/15 \\
	&  $10^{-5}$  &  0.57  & 47.22 &16.26/15 \\
S-3	&  $10^{-4}$  &  6.06 & 46.19   &11.45/15  \\
	&  $10^{-3}$  &  64.01  & 44.91  &  11.71/15\\
	\vspace{1.5mm}
	&   $10^{-2}$  & 1241.48  & 43.88  & 32.32/15 \\
\hline		

	&  $10^{-6}$  &  0.06 & 48.19  & 8.43/14 \\
	&  $10^{-5}$  &  0.61  & 46.94  & 8.77/14\\
S-4	&  $10^{-4}$  &  6.58 &  45.94   & 5.90/14  \\
	&  $10^{-3}$  &  67.44 & 44.74  & 6.08/14\\
	\vspace{1.5mm}
	&   $10^{-2}$ & 1118.08  & 44.12 & 18.23/14 \\

\hline		
	  &  $10^{-6}$ &  0.06 & 48.57 & 5.47/14 \\
	  &  $10^{-5}$ &  0.69 & 47.19  & 5.59/14\\
QS-1 &  $10^{-4}$  & 7.09  & 45.97   & 4.87/14  \\
	  &  $10^{-3}$  & 76 & 44.7   & 5.11/14\\
	\vspace{1.5mm}
	  &   $10^{-2}$  & 1113.37 & 44.26   & 27.64/14 \\	
	
\hline		
	&  $10^{-6}$ &  0.09 & 48.09 & 14.07/14 \\
	&  $10^{-5}$ &  0.94  & 46.78  & 13.94/14\\
QS-2	&  $10^{-4}$  & 9.55 & 45.57  & 14.82/14 \\
	&  $10^{-3}$  &  87   & 44.57   & 6.47/14\\
	\vspace{1.5mm}
	&   $10^{-2}$  &  1343 & 43.34   & 178/14 \\	
	
\hline		
\end{tabular}
\end{adjustbox}
\label{table:jet_power}
\end{table*}

\begin{table*}
\caption{Best-fit  parameters obtained by fitting the local convolution SED  model  involving SSC nad EC processes to  S-1, S-2, S-3, S-4, QS-1 and QS-2 states. 
Row:- 1: particle index before the break energy, 2: particle index after the break energy, 3: bulk Lorentz factor of the emission region, 4: magnetic field in units of $10^{-3}\rm G$,  5: $\rm \xi_{break}$ parameter, which represent break energy   $\rm \gamma_{break}$ through the relation  $\rm \gamma_{break}=\xi_{break}/\sqrt{\mathbb{C}}$, where $\rm \mathbb{C} =1.36\times 10^{-11}\delta B/(1+z)$, 6: equipartition parameter value  7: logarithmic jet power in units of $\rm{erg\,s^{-1}}$,  8:  $\chi^2$/degrees of freedom,  and 9:  Galactic neutral hydrogen column density ($\rm n_H$) in units of $\rm cm^{-2}$.  The subscript and superscript values on parameters are lower and upper values of model parameters respectively obtained through spectral fitting. $--$ implies that the upper or lower bound value on the parameter is not constrained. For each of the flux states, size of emission region and viewing angle are  chosen as $\rm 10^{17} cm $ and 0.1 degree, respectively, $\rm \xi_{min}$ values are chosen  within the range of  $10^{-4}-10^{-3}$,  and $\rm \xi_{max}$ is selected between $1-4$.}
\begin{adjustbox}{width=0.9\textwidth,center=\textwidth}
\begin{tabular}{lcccccccccc}
\bottomrule
Free Parameters & S-1 & S-2  & S-3 & S-4  & QS-1 & QS-2 \\
\bottomrule
\vspace{2.5mm}
$\rm p$ & $2.64_{2.62}^{2.65}$  & $2.35_{2.26}^{2.44}$ &  $2.32_{2.22}^{2.43}$   & $2.31_{2.23}^{2.36}$ & $2.38_{2.32}^{2.46}$ & $2.69_{2.68}^{2.71}$  \\
\vspace{1.5mm}
$\rm q$ & $4.40_{4.22}^{4.64}$ & $3.55_{3.47}^{3.62}$ &  $4.04_{3.94}^{4.16}$  & $3.92_{3.85}^{4.02}$ & $4.23_{4.10}^{4.39}$ & $4.25_{3.98}^{4.54}$ \\
\vspace{1.5mm}
$\Gamma$ & $20.09_{19.74}^{20.81}$ & $20.38_{19.06}^{22.33}$  &   $18.82_{17.98}^{19.79}$    & $18.45_{16.35}^{23.20} $& $14.32_{14.10}^{14.78}$ & $6.01_{5.62}^{6.42}$  \\
\vspace{1.5mm}
$\rm B$ & $0.50_{0.46}^{0.55}$ & $0.72_{0.65}^{0.77}$ &  $0.49_{0.45}^{--}$   &    $0.51_{0.47}^{0.56}$ & $0.50_{0.48}^{0.52}$ & $0.84_{0.70}^{1.02}$  \\

\hline
Fixed parameters  & & & \\ 

\hline

\vspace{1.5mm}
$\rm \xi_{break}$ & $4.25\times 10^{-2}$  &  $1.01\times 10^{-2}$ & $3.78\times 10^{-2}$ & $3.24\times 10^{-2}$ &  $2.92\times 10^{-2}$  & $2.01\times 10^{-2}$ \\
\vspace{1.5mm}
$\rm P_{jet}$ & 45.19 & 45.10 &  45.42  &  45.25 & 44.95 & 44.40\\
$\eta$ &1.00& 1.17 & 1.24 & 0.90 & 1.20  & 1.32   \\

\vspace{1.5mm}
$\rm n_H$ & $2.8\times 10^{21}$ & $2.8\times 10^{21}$ & $2.0\times 10^{21}$ & $2.2\times 10^{21}$ &  $2.8\times 10^{21}$ & $2.8\times 10^{21}$   \\

\vspace{1.5mm}
$\rm \chi^2/dof$ & 4.62/14 & 7.41/14 & 14.63/15 & 5.91/14 &  5.01/14  & 5.02/14   \\

\hline
\end{tabular}
\end{adjustbox}
\label{table:sed}
\end{table*}

\section{Summary and Discussion}\label{sec:discus}

The \emph{Fermi} light curve of BL\,Lac during  period MJD 59000--59943 shows the source was in active state for long  time. During this period, $\gamma$-ray light curve revealed presence  of multiple flaring components, a maximum daily averaged  $\gamma$-ray flux of $\rm (1.74 \pm 0.09) \times 10^{-5}  ph\, cm^{-2} s^{-1}$ is observed on MJD 59868.5 .
 This is the highest one day binned $\gamma$-ray flux detected from source. A shortest flux doubling timescale of $t_{var} = 0.40$ d is observed during the time interval 59022- 59023.
The $\gamma$-ray light curve require a series of exponentials to reproduce the profile shape. 
We noted that among dominant components, five components are moderately asymmetric, one component is asymmetric and six components are symmetric. The symmetry in the  flare profile can be due to the light travel time effects,  while the asymmetry in the flare profile could be attributed to strengthening and weakening of acceleration process. A slow rise in the asymmetric  flare possibly indicates  acceleration of particles to higher energy, while fast decay may be associated with the rapid  energy loss of high energy particles. 
Moreover, a usual harder when brighter trend is observed  in the $\gamma$-ray light curve. In case of BL\,Lac, the $\gamma$-ray spectrum lies near the peak of IC component, therefore the spectral hardening during the flaring indicates the shift in Compton peak towards high energy.
The spectral hardening  and shift in SED peak energy of IC component during high flux states had been observed in 3C 279 by \citet{2019MNRAS.484.3168S}.

We calculated PSD of  $\gamma$-ray light curve in order to obtain an insight into the physical processes causing large variability in BL\,Lac object. We found that the  PSD is a  power law with an  index   $\sim 1$, which suggest for a flicker  noise type process. Similar results have also been reported by \citet{2014ApJ...786..143S, 2020ApJ...891..120B}.
In their work,  PSD  of  $\gamma$-ray light curves of sample of blazars were found to be consistent with a  power law where majority of  the indexes were near to 1.0. The flicker noise is halfway between random walk (index $= 2$) and white noise (index $= 0$), 
 it has the property of maintaining shape over several orders of frequencies up to arbitrarily low values. Therefore,  observation of such feature in the light curve implies long memory process is at work. 
 In other words, it implies that the  shorter and longer timescale variations are coupled together or equivalently the underlying processes should be multiplicative.
 In blazars, the jet emission can possess memory of the events occurring at the accretion disk, especially the disk modulations  and thus indicates for disk-jet connection.

We also characterised the variability of $\gamma$-ray light curve of BL\,Lac object by obtaining a correlation between the flux and r.m.s. The r.m.s-flux plot shows a linear trend, which  has been noted in several other astrophysical systems like black hole binaries  and  Active Galaxies \citep{2001MNRAS.323L..26U,2003MNRAS.345.1271V, 2005MNRAS.359..345U}. The linear r.m.s-flux relation implies that short and long variability time-scales are coupled together in multiplicative way \citep{2005MNRAS.359..345U}. It also rules out shot-noise models where the different time-scales of  variability are combined additively. In addition to r.m.s-flux relation, we checked the flux distribution of the $\gamma$-ray light curve with the skewness, AD test and histogram fitting. All these tests reject the normality of the flux distribution,  instead suggest that the flux distribution is lognormal. 
Observation of lognormal distribution  imply that the underlying emission process responsible for the variability is multiplicative rather than additive.  
These features  are  mostly believed to be result of perturbations  in the accretion disk \citep{2005MNRAS.359..345U,  2008bves.confE..14M,  1997MNRAS.292..679L}. In the fluctuating accretion model, the perturbation in the mass accretion rate can propagate inwards and  accure in multiplicative way in the inner regions of the disc, thereby producing the multiplicative emission.  
In case of blazar, the emission mainly comes from the jet, hence  possible realisation for the observation of lognormal behavior could be that the disk fluctuations are imprinted on the jet emission \citep{2009A&A...503..797G, 2010LNP...794..203M, 2018RAA....18..141S}.  On the other hand, the minute time scale variation observed in the $\gamma$-ray light curves \citep{1996Natur.383..319G,  2007ApJ...664L..71A}  implies that the jet emission should be independent of accretion disc fluctuations \citep{2012MNRAS.420..604N}. In such cases, the lognormal distribution in flux can be explained by the linear Gaussian perturbations in  particle acceleration time-scales \citep{2018MNRAS.480L.116S}.  Moreover, the fluctuations in the escape time scales of the electrons would produce flux distribution shapes other than Gaussian or lognormal. In addition,  \citet{2012A&A...548A.123B} have shown that  the additive shot noise-model can also produce a lognormal flux distribution, such as the Doppler boosting of emission from a large number of randomly oriented mini-jets results in a flux distribution with features similar to that of a lognormal distribution.

We examined the broadband spectral characteristics of BL\,Lac by choosing the time intervals for which the  simultaneous observations are available in $\gamma$-ray, X-ray and UV/optical bands. The convolved one zone leptonic model suggests that the underlying particle energy distribution responsible for the broadband emission is more likely to be a BPL  distribution.  Therefore,  the broad band SEDs in different flux states are statistically modelled  by using the BPL electron energy distribution which undergo synchrotron and SSC loss.  The statistical fit is carried by keeping p, q, $\Gamma$, and  $\rm B$ as free parameters, while other parameter are kept fixed to the typical values required by the particular flux state.
We showed that the optimal $\rm P_{jet}$ in different flux states  (see Table \ref{table:jet_power}) are obtained by choosing the  $\rm \xi_{min}$ within the range $10^{-4}$ to $10^{-3}$.  Under the conditions of  equipartition and minimum jet power, the best fit parameters (shown in Table \ref{table:sed}) implies that the increase in flux from low to high flux state is  associated with the increase in $\Gamma$. Additionaly,  the particle spectral indices becomes harder in the high flux state. 

 The SED of BL Lac has been the subject of multiple modeling attempts in the past. 
During the era of the EGRET satellite, it had been  observed that SED modeling of  the high flux states above 100 MeV requires the inclusion of external seed photons for Compton scattering. For example,  \citet{2000AJ....119..469B} showed that SSC and EC emission is required  to yield an acceptable fit to the broadband spectrum. They showed that in BL Lac, unlike other BL Lac objects, the broad emission line region plays an important role for the high energy emission.   Using the broadband emission model  of \citet{2002A&A...383..763R},  \citet{2007ApJ...666L..17A} showed that one zone SSC model can explain the broad band emission upto VHE energies during  relatively weak $\gamma-$ray emission,  while the strong $\gamma-$ray emission during the 1997 flare requires SSC as well as EC emission components for the broadband emission.
  Additionally, \citet{2011ApJ...730..101A} showed that the SED may be described by a single zone or two zone SSC model, but a hybrid SSC plus EC model is preferred based on the observed variability.  
In this work, we  constrained the underlying particle distribution  using the $\chi^2$ test. We noted that synchrotron and SSC emission provided a  reasonable fit to all the flux states considered in our analysis. However, the equipartition values obtained  are much larger than unity. Therefore we  modelled the  broad band emission  from BL\,Lac by considering the EC emission alongside synchrotron and SSC emissions. These three emission processes ensure the equipartition between the magnetic energy density and the particle energy density.
The best fit broken power law indices obtained show that index after break energy are steeper than what would be expected from the  cooling break.  The results thus rules out the broken power law spectrum as originating from a radiative cooling.  The exact explanation for this steep spectrum remains unclear. It could potentially result from  multiple acceleration processes or energy dependence of diffusion time scales. For example, a steep  electron spectrum  may possibly  be interms of an arbitrary energy dependence of the diffusion coefficient. \citet{2007A&A...465..695Z} studied the energy spectra of shock accelerated electrons and their associated radiation by considering an arbitrary energy-dependence of the diffusion coefficient. The authors show that in case of  Bohm diffusion, the spectral cutoff takes on a sub-exponential form/steeper cutoff at high energies. Within this framework,  the electron spectrum exhibits a significantly faster decay rate  than the broken power law observed in our work.
 Alternatively,  \citet{2008MNRAS.388L..49S} proposed a two-zone model where  broken power law  injection into the cooling region undergoing synchrotron losses. 
 introduces an additional break  in the electron spectrum with indices $\rm p+1$ and $\rm p+2$ where p is the index of spectrum  before the break energy $\rm \gamma_b$.  Nonetheless, the presence of this additional break in the spectrum within our work can not be confirmed due to the energy gap between the observed optical/UV and X-ray spectra.

The  MLC reveal  that the BL\,Lac object exhibits  a strong correlated variability at  optical/UV, X-rays and $\gamma$-ray bands.  Analysis, particularly using the ZDCF, shows a positive correlation among the emissions in different bands without any significant time lag between them.
 This implies that single emission region and  same electron population are responsible for the emission in different energy bands during a particular flux state. Moreover, the light curves show large  amplitude of variability  in X-ray flux compared to the  variability in optical/UV and $\gamma$-ray band. A similar result was reported by \citet{2021MNRAS.507.5602P}. 
 This trend may be attributed  to  the shape of the broad band SED.  The broadband SED of BL\,Lac source during high flux states (see Figures \ref{fig:s12} and \ref{fig:s34}) shows that the X-ray spectrum  lie after the break energy. This implies that the X-ray emission are due to the high energy electrons, while the $\gamma-ray$ and optical/UV emission are comparatively due to low energy electrons. Since the high energy electrons cool faster, theoretically one expects larger amplitude variability in flux at X-ray band compared to the $\gamma$-ray and optical/UV band.

\section{ACKNOWLEDGEMENTS}
The research work is supported by the Department of Science and Technology, Government of India, under the INSPIRE Faculty grant (DST/INSPIRE/04/2020/002319). The author acknowledges the support of Prof. Ranjeev Misra in obtaining the Convolution codes used in this work. This research has made use of $\gamma$-ray data from Fermi Science Support Center (FSSC). The work has also used the \emph{Swift} Data from the High Energy Astrophysics Science Archive Research Center (HEASARC), at NASA's Goddard Space Flight Center. The author expresses gratitude to the anonymous referee for providing valuable and insightful comments.

\section{Data availability}
The data and the model used in this article will be shared on reasonable request to the corresponding author, Zahir Shah (email: shahzahir@cukashmir.ac.in or shahzahir4@gmail.com).

\bibliographystyle{mnras}
\bibliography{references}

\bsp	
\label{lastpage}
\end{document}